\newif\ifARXIV
\ARXIVtrue

\newif\ifDEBUG
\DEBUGfalse

\newif\ifNOBOXES
\NOBOXESfalse

\newif\ifSPACEHACK
\SPACEHACKfalse

\newif\ifANONYMOUS
\ANONYMOUStrue

\ifARXIV
  \documentclass[letterpaper,twocolumn,10pt]{article}

  \usepackage{usenix2019_v3}
\else
  \ifANONYMOUS
    \documentclass[sigconf,anonymous,review]{acmart}
    \setcopyright{none} 
    \renewcommand\footnotetextcopyrightpermission[1]{}
    \acmConference[ICSE '26]{48th International Conference on Software Engineering}{April 12--18, 2026}{Rio de Janeiro, Brazil}
  \else
  \fi
\fi

\usepackage{amsmath,amssymb,amsfonts}

\usepackage{algorithmic}
\usepackage{graphicx}
\usepackage{textcomp}
\usepackage{caption}
\usepackage{xcolor}
\usepackage{booktabs} 
\usepackage{xspace} 
\usepackage[normalem]{ulem}
\usepackage{makecell}
\usepackage{listings}
\usepackage{listing}
\usepackage{tcolorbox}
\usepackage{enumitem}
\usepackage{siunitx}
\usepackage[vskip=1em,font=itshape,leftmargin=2em,rightmargin=2em]{quoting}
\usepackage{tabularx}
\usepackage{hyperref}
\hypersetup{
  breaklinks=true,
  colorlinks=true,
  urlcolor=blue
}
\usepackage{lscape}
\usepackage{afterpage}

\usepackage{makecell}
\usepackage{multirow}
\usepackage{booktabs}
\usepackage[T1]{fontenc}

\usepackage{setspace}

\usepackage{tikz}

\usepackage{wrapfig}

\usepackage{enumitem}
\usepackage{xurl}

\usepackage{lipsum}

\usepackage{soul}

\ifDEBUG
    \newcommand{\JD}[1]{\textcolor{purple}{[JD:#1]}}
    \newcommand{\AG}[1]{\textcolor{olive}{[AG:#1]}}
    \newcommand{\WJ}[1]{\textcolor{teal}{[WJ:#1]}}
    \newcommand{\GKT}[1]{\textcolor{brown}{[GKT:#1]}}
    
    \newcommand{\NV}[1]{\textcolor{red}{[NV: #1]}}
    \newcommand{\CC}[1]{\textcolor{violet}{[CC: #1]}}
    \newcommand{\Andreas}[1]{\textcolor{cyan}{[ADK: #1]}}
     \newcommand{\YT}[1]{\textcolor{red}{[YT: #1]}}
     
     \newcommand{\PA}[1]{\textcolor{violet}{[PA: #1]}}
    \newcommand{\BC}[1]{\textcolor{blue}{[BC: #1]}}

    \newcommand{\TODO}[1]{\hl{#1}}
\else
    \newcommand{\JD}[1]{}
    \newcommand{\AG}[1]{}
    \newcommand{\WJ}[1]{}
    \newcommand{\GKT}[1]{}
    \newcommand{\NS}[1]{}
    \newcommand{\NV}[1]{}
    \newcommand{\CC}[1]{}
    \newcommand{\JY}[1]{}
    \newcommand{\YT}[1]{}
    \newcommand{\PA}[1]{}
    \newcommand{\BC}[1]{}

    \newcommand{\TODO}[1]{#1}
\fi

    \usepackage{etoolbox}
    \makeatletter
    \patchcmd{\ttlh@hang}{\parindent\z@}{\parindent\z@\leavevmode}{}{}
    \patchcmd{\ttlh@hang}{\noindent}{}{}{}
    \makeatother

\newcommand{\myparagraph}[1]{\vspace{0.10cm}\noindent\hspace{0.25cm}\uline{\textit{#1}} \noindent{}}

\ifSPACEHACK
    \titlespacing*\section{0pt}{5pt plus 1pt minus 1pt}{3pt plus 1pt minus 1pt}
    \titlespacing*\subsection{0pt}{4pt plus 1.5pt minus 1.5pt}{4pt plus 1.5pt minus 1.5pt}
    \titlespacing*\subsubsection{0pt}{3pt plus 1pt minus 1pt}{3pt plus 1.5pt minus 1.5pt}
    \titlespacing*\paragraph{0pt}{1pt plus 1.5pt minus 1.5pt}{2pt plus 1.5pt minus 1.5pt}
\fi

\usepackage{balance} 
\usepackage[font=small,labelfont=bf]{caption}
\usepackage{subcaption}
\usepackage[bottom]{footmisc}
\makeatletter
\def\cl@chapter{}
\makeatother
\usepackage{cleveref}

\crefformat{section}{\S#2#1#3}
\crefname{figure}{Figure}{Figures}
\crefname{appendix}{Appendix}{Appendices}
\crefname{table}{Table}{Tables}
\crefname{algorithm}{Algorithm}{Algorithms}
\crefname{listing}{Listing}{Listings}
\crefname{theorem}{Theorem}{Theorems}
\crefname{thm}{Theorem}{Theorems}
\crefname{lemma}{Lemma}{Lemmata}
\crefname{equation}{Eqt.}{Eqts.}
\crefformat{Grammar}{Grammar #1}

\usepackage{seqsplit}
\usepackage{enumitem}
\usepackage{booktabs}
\usepackage[ruled,vlined]{algorithm2e}

\newcommand{\ie}{\textit{i.e.,} }
\newcommand{\eg}{\textit{e.g.,} }
\newcommand{\etal}{\textit{et al.}\xspace}
\newcommand{\etals}{\textit{et al.'s}\xspace}

\newcommand{\vs}{{\em vs.}\xspace}

\newenvironment{RQList}{
   \setlength{\topsep}{0pt}
   \setlength{\partopsep}{0pt}
   \setlength{\parskip}{0pt}
   \begin{description}[style=unboxed]
   \setlength{\leftmargin}{1in}
   \setlength{\parsep}{0pt}
   \setlength{\parskip}{0pt}
   \setlength{\itemsep}{0pt}
   }
   {\end{description}}
   
\definecolor{delim}{RGB}{20,105,176}
\definecolor{numb}{RGB}{106, 109, 32}
\definecolor{string}{rgb}{0.64,0.08,0.08}

\lstdefinelanguage{json}{
    showspaces=false,
    showtabs=false,
    breaklines=true,
    postbreak=\raisebox{0ex}[0ex][0ex]{\ensuremath{\color{gray}\hookrightarrow\space}},
    breakatwhitespace=true,
    basicstyle=\ttfamily\small,
    upquote=true,
    morestring=[b]",
    stringstyle=\color{string},
    literate=
     *{0}{{{\color{numb}0}}}{1}
      {1}{{{\color{numb}1}}}{1}
      {2}{{{\color{numb}2}}}{1}
      {3}{{{\color{numb}3}}}{1}
      {4}{{{\color{numb}4}}}{1}
      {5}{{{\color{numb}5}}}{1}
      {6}{{{\color{numb}6}}}{1}
      {7}{{{\color{numb}7}}}{1}
      {8}{{{\color{numb}8}}}{1}
      {9}{{{\color{numb}9}}}{1}
      {\{}{{{\color{delim}{\{}}}}{1}
      {\}}{{{\color{delim}{\}}}}}{1}
      {[}{{{\color{delim}{[}}}}{1}
      {]}{{{\color{delim}{]}}}}{1},
}

\newcommand{\StepLabel}[1]{%
  \textbf{Part~\textcolor{blue}{\textcircled{\small#1}}}%
}
\newcommand{\StepRef}[1]{%
  \hyperref[sec:SystemDesign-Step#1]{\StepLabel{#1}}%
}
\newcommand{\tool}{\texttt{ConfuGuard}\xspace}
\newcommand{\FPReducedRate}{52\%\xspace}
\newcommand{\NumRemovedPkgs}{1,003\xspace}
\newcommand{\NumReviewedThreats}{630\xspace}

\newcommand{\NumNeupaneDB}{1,840\xspace}
\newcommand{\NumConfuDB}{2,361\xspace} 
\newcommand{\NumNeupaneDBConfuguardEvaluated}{1,819\xspace}
\newcommand{\NumConfuDBConfuguardEvaluated}{2,034\xspace} 

\newcommand{\NumTypomindData}{1,239\xspace}

\newcommand{\TypomindFPNum}{640,482\xspace}
\newcommand{\SampledTypomindFPNum}{626\xspace}
\newcommand{\SampledAvailableTypomindFPNum}{601\xspace}
\newcommand{\InterRaterAgreement}{0.6\xspace}

\newcommand{\FPBenignPkgs}{480\xspace}
\newcommand{\FPStealthyPkgs}{121\xspace}

\newcommand{\FPBenignPkgsPercent}{79.9\%\xspace}
\newcommand{\FPStealthyPkgsPercent}{20.1\%\xspace}
\newcommand{\TypomindFPRate}{80\%\xspace}
\newcommand{\ConfuguardFPRate}{28\%\xspace}

\newcommand{\FPPrelimDataConfidenceLevel}{99\%\xspace}
\newcommand{\FPPrelimDataMarginError}{5\%\xspace}

\newcommand{\NumAllGolangPkgs}{175K\xspace}
\newcommand{\NumAllMavenPkgs}{503K\xspace}
\newcommand{\NumAllPyPIPkgs}
{619K\xspace}
\newcommand{\NumAllRubyPkgs}{212K\xspace}
\newcommand{\NumAllHFPkgs}{1.1M\xspace}
\newcommand{\NumAllNPMPkgs}{5.1M\xspace}
\newcommand{\NumAllNugetPkgs}{430K\xspace}

\newcommand{\NumAllPopGolangPkgs}{20,713\xspace}
\newcommand{\NumAllPopMavenPkgs}{27,831\xspace}
\newcommand{\NumAllPopPyPIPkgs}
{9,525\xspace}
\newcommand{\NumAllPopRubyPkgs}{60,695\xspace}
\newcommand{\NumAllPopHFPkgs}{13,252\xspace}
\newcommand{\NumAllPopNPMPkgs}{36,333\xspace}

\newcommand{\NumProdUnreviewedTypos}{50,069\xspace}

\newcommand{\NumProdReviewedTPTypos}{630\xspace}

\begin{document}






\title{\tool: Using Metadata to Detect Active and Stealthy Package Confusion Attacks Accurately and at Scale}
\ifARXIV
    \author{
  Wenxin Jiang \\
  Purdue University\\
  \and
  Berk Çakar \\
  Purdue University \\
  \and
  Mikola Lysenko \\
  Socket, Inc. \\
  \and
  James C. Davis \\
  Purdue University \\
}

\else
    \author{Wenxin Jiang}
    \affiliation{%
      \institution{Electrical and Computer Engineering\\Purdue University}
      \city{West Lafayette}
      \state{IN}
      \country{USA}}
    \email{jiang784@purdue.edu}
    
    \author{Berk Çakar}
    \affiliation{%
      \institution{Electrical and Computer Engineering\\Purdue University}
      \city{West Lafayette}
      \state{IN}
      \country{USA}}
    \email{bcakar@purdue.edu}
    
    \author{Mikola Lysenko}
    \affiliation{%
      \institution{Socket, Inc.}
      \city{San Francisco}
      \state{CA}
      \country{USA}}
    \email{mik@socket.dev}
    
    \author{James C. Davis}
    \affiliation{%
      \institution{Electrical and Computer Engineering\\Purdue University}
      \city{West Lafayette}
      \state{IN}
      \country{USA}
    }
    \email{davisjam@purdue.edu}
\fi

\maketitle
\begin{abstract}
Package confusion attacks such as typosquatting
threaten software supply chains.
Attackers make packages with names that syntactically or semantically resemble legitimate ones, tricking engineers into installing malware.
While prior work has developed defenses against package confusions in some software package registries, notably NPM, PyPI, and RubyGems, gaps remain:
  high false-positive rates,
  generalization to more software package ecosystems,
  and
  insights from real-world deployment.

In this work, we introduce \tool, a state-of-art detector for 
  package confusion threats.
  We begin by presenting the first empirical analysis of benign signals derived from prior package confusion data, uncovering their threat patterns, engineering practices, and measurable attributes.
Advancing existing detectors, we
  leverage package metadata to distinguish benign packages,
  and
  extend support from three up to seven software package registries.
Our approach significantly reduces false positive rates (from \TypomindFPRate to \ConfuguardFPRate), at the cost of an additional \(14s\) average
latency 
to filter out benign packages by analyzing the package metadata.
\tool is used in production at our industry partner, whose analysts have already confirmed \NumReviewedThreats real attacks detected by \tool. 

\end{abstract}
\section{Introduction}

\begin{figure}[t]
    \centering
    \includegraphics[width=0.98\linewidth, trim=10 10 7 15, clip]{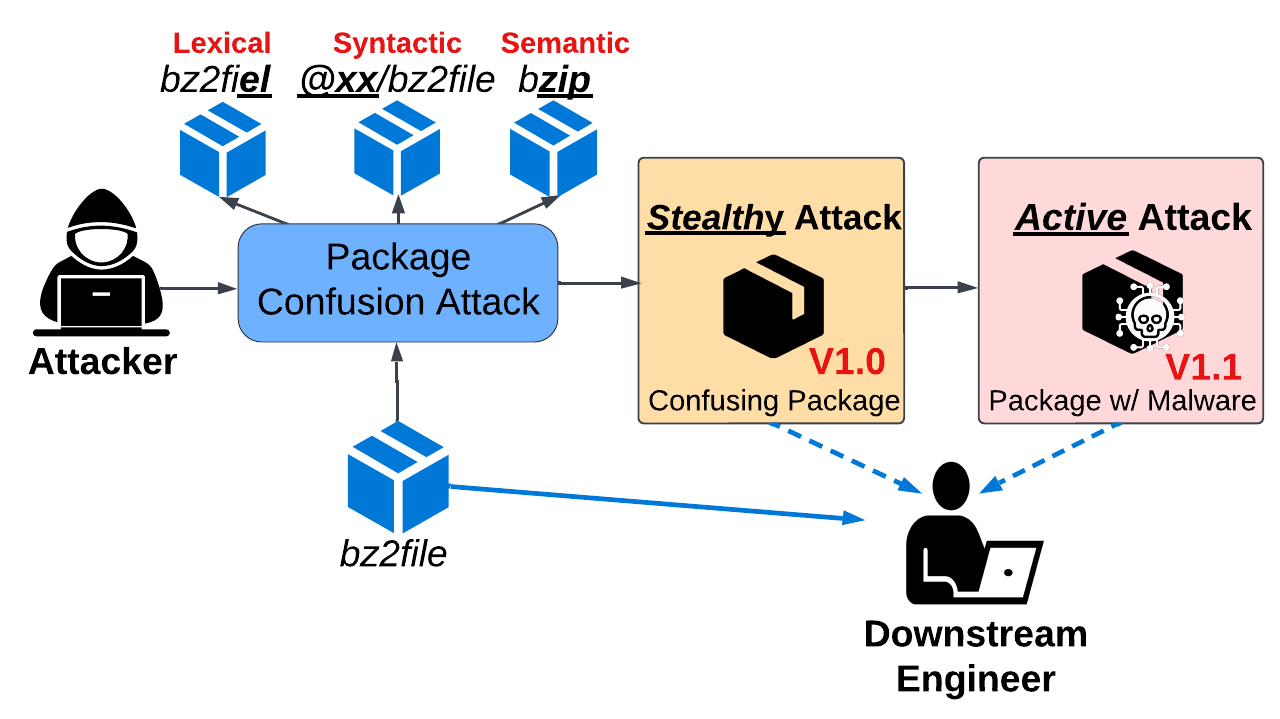}
    \caption{
    Threat model illustrating common active and stealth package confusion attacks.
    Attackers spoof a legitimate package such as \texttt{bz2file} with lexical typos (\texttt{bz2fiel}), syntactic namespace manipulations in scoped registries (\texttt{@xx/bz2file}), and semantic substitutions (\texttt{bzip}).
    Attackers release these look-alike packages as \textit{active} attacks (malicious from the first version) or \textit{stealth} attacks that are benign until a later update injects malware.
    Engineers may be tricked into depending on the impostor directly or as a transitive dependency. 
    }
   \label{fig:ThreatModel}
\end{figure}

Software package registries (SPRs) are indispensable in modern development.
They help engineers develop applications by integrating open-source packages, reducing costs and accelerating product cycles~\cite{raymond_cathedral_1999,SotoValero2019EmergenceofSWDiversityinMavenCentral, Wittern2016JSPackageEcosystem_new}.
SPRs are used by industry and governments~\cite{18fopensourcepolicy, synopsys2024ossra}, including in AI and safety-critical applications~\cite{Jiang2022PTMReuse, stenbit2003OSSinDOD}.
The resulting software supply chain is an attractive target --- if an adversary can cause an engineer to install a malicious package, the attacker may gain access to downstream users' resources.
One of their tactics is \textit{package confusion attacks}~\cite{kaplan_survey_2021, neupane2023beyondTyposquatting}: creating malicious packages with misleading names~\cite{gu2023investigatingPackageRelatedSecurityThreatsinSoftwareRegistries, Ladisa2023TaxonomyofAttacksonOSSSC, okafor_sok_2022}. 
\cref{fig:ThreatModel} shows the attack model, covering naming tactics (lexical, syntactic, and semantic confusion) and attack timing (active \vs stealthy).

Researchers have proposed detectors to mitigate package confusion attacks.
Early solutions detected packages with lexically similar names~\cite{vu2020typosquatting, taylor2020spellboundDefendingAgainstPackageTyposquatting,sern2021typoswype}.
State-of-the-art approaches have begun exploring semantic similarity~\cite{neupane2023beyondTyposquatting}.
However, as criticized by Ohm \etal~\cite{ohm2023SoKPracticalDetectionofSSCAttacks},
prior work did not emphasize low false positive rates, which are needed for real-world use~\cite{ohm2023SoKPracticalDetectionofSSCAttacks}.
We note also that
  prior work focuses on limited SPRs, notably NPM and PyPI;
  and
  that the literature lacks experience reports from production deployment.

In this study we propose a novel system to detect package confusion attacks. 
To reduce false positive rates, we examine the practices of package confusion attackers and propose signals, based on package metadata, to distinguish benign packages from genuine (though perhaps stealthy) threats.
To generalize prior work to more SPRs, we analyze naming practices and confusion attacks in seven SPRs, and identify novel aspects of these SPRs that expand the attack surface for package confusion.
We also describe an embedding-based name comparison method and a novel nearest-neighbor search algorithm that improve the accuracy and efficiency of detection. 

We integrated these techniques in the \tool system.
We initially evaluated \tool on previous package confusion attack data\-sets, and iteratively enhanced it using recent real-world package confusion reports flagged and triaged by our industry partner's security analysts.
We report that \tool achieves a false positive rate of \ConfuguardFPRate, an improvement of \FPReducedRate over the state of the art.
\tool has average latency of 6.8 seconds per package on a modern server.
Our industry partner has already used \tool in production to identify \NumReviewedThreats confirmed package confusion attacks. 

To summarize our contributions:
\begin{itemize}[leftmargin=1.5em, itemsep=0.1ex]
    \item 
    We refine the definition of package confusion attacks to encompass both active and stealthy threats.
    \item We generalize prior analyses to several new SPRs, noting the effect of package naming hierarchies on confusion attack tactics.
    \item We propose novel algorithms to improve the accuracy and efficiency of detecting package confusion attacks.
    \item We open-source \texttt{ConfuDB}, our production dataset containing \NumConfuDB package confusion threats triaged by security analysts.
    \item We share lessons learned from using \tool in production.
\end{itemize}

\noindent
{\ul{\textbf{Significance:}}}
Package confusion attacks are common yet challenging to detect.
This paper introduces \tool, a best-in-class detection system for package confusion detection,
designed through analysis of previous attacks and insights from ongoing deployment.
\tool is used in production by our industry partner. 
\section{Background and Related Work} 
Here, we provide background on SPRs and their package naming conventions (\cref{sec:background-SPR-PkgNaming}), review the taxonomy of package confusion attacks (\cref{sec:background-Taxonomy}), and outline existing defense mechanisms (\cref{sec:background-Defenses}).

\subsection{SPRs and Package Naming}
\label{sec:background-SPR-PkgNaming}

Much of modern software development depends on community ecosystems of libraries and frameworks that facilitate software reuse~\cite{Wittern2016JSPackageEcosystem_new, Zimmermann2019SecurityThreatsinNPMEcosystem}.
We refer to these reusable software artifacts as \textit{software packages}.
These packages are distributed by \textit{software package registries~(SPRs)}~\cite{SotoValero2019EmergenceofSWDiversityinMavenCentral, schorlemmer2024signing}.
\cref{tab:registry-overview} shows the SPRs for seven popular ecosystems comprising over 8 million software packages.
Some SPRs restrict package publication to individual accounts, so the publisher's identifier does not appear in the package name. Others allow grouping packages into namespaces or organizations, which are incorporated into the package name.
We call the former package naming structure ``flat'' and the latter ``hierarchical''~\cite{balakrishnan2004layeredNamingArchitecture}.

\begin{table}[ht]
\footnotesize
\setlength{\tabcolsep}{2pt}
\caption{
SPRs analyzed in this study, including their programming ecosystem, package naming structure, and presence in prior work on package confusion detection.
Registry sizes are as of Jan. 2025.
An example flat naming pattern is PyPI's \texttt{requests} module.
An example hierarchical name is Hugging Face's \texttt{google-bert/bert-base-uncased}.
}
\label{tab:registry-overview}
\footnotesize
\centering
\resizebox{0.95\linewidth}{!}{%
\begin{tabular}{lcccc}
\toprule
\textbf{Registry\ (\#\ pkgs)} & \textbf{Ecosystem} & \textbf{Name\ Structure} & \textbf{Prior\ Work} \\
\midrule
PyPI (\NumAllPyPIPkgs)
& Python
& Flat
& \cite{vu2020typosquatting, neupane2023beyondTyposquatting}
\\

RubyGems (\NumAllRubyPkgs)
& Ruby
& Flat
& \cite{neupane2023beyondTyposquatting}
\\

Maven (\NumAllMavenPkgs)
& Java
& Hierarchical
& N/A
\\

Golang (\NumAllGolangPkgs)
& Go
& Hierarchical
& N/A
\\

Hugging Face (\NumAllHFPkgs)
& AI Models
& Hierarchical
& \cite{jiang2023PTMNaming}
\\

NPM (\NumAllNPMPkgs)
& JavaScript
& Both
& \cite{taylor2020spellboundDefendingAgainstPackageTyposquatting, neupane2023beyondTyposquatting}
\\

NuGet (\NumAllNugetPkgs)
& .NET
& Both
& N/A
\\

\bottomrule
\end{tabular}
}

\end{table}


SPRs typically act as infrastructure and leave it to the engineering community to establish norms for interaction.
One aspect of their ``hands-off'' approach is that SPRs impose almost no constraints on package names --- they define an allowed character set and forbid duplicate names~\cite{NPMPackageName, rubygemsNaming, golangPackageNames, pep423, mavenNamingConventions, jfrogHuggingFaceNaming}.
This flexibility introduces two related challenges for engineers.
First, based on the application domain (\eg traditional software libraries \vs machine learning model packages), engineers across different SPRs have adopted different naming conventions, which can enhance readability, maintainability, and collaboration~\cite{butler2010exploring, jiang2023PTMNaming}, but which may also confuse newcomers.
Second, as we detail next, SPRs' permissive naming practices enable adversaries to choose package names to deliberately confuse engineers.

\subsection{Package Confusion Attacks and Taxonomy}
\label{sec:background-Taxonomy}
\textit{Package confusion}
is an attack in which adversaries seek to persuade engineers to install malicious packages~\cite{neupane2023beyondTyposquatting}.\footnote{Package confusion is the SPR-specific manifestation of the general naming confusion attack, which affects all IT endeavors including DNS (domain names)~\cite{coyle2023chrome,ulikowski2025dnstwist, gabrilovich2002homograph}, BNS (Blockchain names)~\cite{muzammil2024BlockchainTyposquatting}, and container images (image names)~\cite{liu2022exploringContainerRegistryTyposquatting, datadoghqWhoami}.}
Attackers typically choose a package name similar to a legitimate one that meets the engineer's needs~\cite{neupane2023beyondTyposquatting}.
Attacks may be multi-pronged, \eg using advertising campaigns to increase their chances of fooling engineers~\cite{sharma2024SonatypePyPIMalwarePytoileur,he2024FakeStars}.
Package confusion attacks are an ongoing concern, with hundreds of packages detected by researchers in Socket~\cite{socketMaliciousNPMPackage}, ReversingLabs~\cite{reversinglabs_r77_typosquatting}, Orca~\cite{orcaDependencyConfusion}, and others~\cite{birsan2021dependencyconfusion}.



\cref{tab:typosquat_taxonomy} summarizes the state-of-the-art taxonomy of package confusion techniques, based on Neupane \etal~\cite{neupane2023beyondTyposquatting}.
The variations in naming structure in different SPRs also affect the nature of package confusion by SPR.
In SPRs with a flat naming structure, confusion attacks look like the top portion of the table.
In hierarchical SPRs, attackers have a larger naming surface to exploit, as illustrated in the bottom portion of the table.

{
\renewcommand{\arraystretch}{0.5}
\tiny
\begin{table}[ht]
\centering
\caption{
Common package confusion techniques with examples.
The top section presents some of the taxonomy proposed by Neupane \etal~\cite{neupane2023beyondTyposquatting}.
We extend this taxonomy by adding three new patterns (bottom; see~\cref{sec:ProblemState-CaseStudy1-TP}) that generalize confusion strategies observed in additional SPRs.
The examples for those new patterns are drawn from package confusion threats flagged by \tool.
}
\label{tab:typosquat_taxonomy}
\footnotesize
\begin{tabular}{p{0.17\textwidth}p{0.26\textwidth}}
\toprule
\textbf{Technique} & \textbf{Example (\textit{basis $\to$ attack})} \\
\midrule
1-step Levenshtein Distance     & \texttt{crypt\textbf{\ul{o}}} $\to$ \texttt{crypt} \\
\\
Sequence reordering      & \texttt{python-nmap} $\to$ \texttt{\textbf{\ul{nmap-python}}} \\
\\
Scope confusion  & \texttt{\textbf{\ul{@}}cicada\textbf{\ul{/}}render} $\to$ \texttt{cicada\textbf{\ul{-}}render} \\
\\
Semantic substitution  & \texttt{b\textbf{\ul{z2file}}} $\to$ \texttt{b\textbf{\ul{zip}}} \\
\\
Alternate spelling  & \texttt{colorama} $\to$ \texttt{colo\textbf{\ul{u}}rama} \\
\midrule
\textbf{Impersonation Squatting}  & \texttt{\textbf{\ul{meta}}-llama/Llama-2-7b-chat-hf} $\to$ \newline\texttt{\textbf{\ul{facebook}}-llama/Llama-2-7b-chat-hf}\\
\\
\textbf{Compound Squatting}  & \texttt{@typescript\textbf{\ul{-}}eslint/eslint\textbf{\ul{-plugin}}} $\to$
\newline \texttt{@typescript\textbf{\ul{\_}}eslint\textbf{\ul{er}}/eslint} \\
\\
\textbf{Domain Confusion} (Golang)  & \texttt{\ul{\textbf{github.com}}/prometheus/prometheus} $\to$
\newline \texttt{\textbf{\ul{git.luolix.top}}/prometheus/prometheus} \\
\bottomrule
\end{tabular}
\vspace{-5mm}
\end{table}
}

\subsection{Defenses Against Package Confusion}
\label{sec:background-Defenses}

Package confusion attacks can be mitigated at many points in the reuse process, \eg by scanning for malware in registries~\cite{vu_lastpymile_2021} or during dependency installation~\cite{latendresse2022PruductionDependenciesinNPM} or by sandboxing dependencies at runtime~\cite{vasilakis_breakapp_2018,ferreira_containing_2021_new,amusuo2025ztd}.
Such techniques provide good security properties, but are computationally expensive.
Our approach is part of a class of cheaper techniques that leverage the most pertinent attribute of a package confusion attack: the similarity of the name to that of another package.
We describe the literature in this vein.

\subsubsection{Approaches} \label{subsubsec:approaches}

In the academic literature, the earliest name-oriented defenses were purely lexical:
Taylor \etal~\cite{taylor2020spellboundDefendingAgainstPackageTyposquatting} and Vu \etal~\cite{vu2020typosquatting}
measured similarity using Levenshtein distance~\cite{levenshtein1966binary} to identify PyPI packages with minor textual alterations from one another.
To detect confusion attacks based on name semantics (\eg \texttt{bz2file} \vs \texttt{bzip}), Neupane \etal~\cite{neupane2023beyondTyposquatting} applied FastText embeddings~\cite{bojanowski2017FasttextEmbedding} to enable an abstract comparison of package names. 

Recognizing the threat of package confusion attacks, industry has also offered some defenses.
NPM and PyPI both use lexical distances to flag packages or otherwise reduce user errors~\cite{NPM_threats_mitigations,psf_acceptable_use_policy,zornstein2023stealthy}.
Some other companies also offer relevant security tools.
For example, Stacklok combines Levenshtein distance with repository and author activity metrics to assign a risk score to suspicious packages~\cite{stacklok_typosquatting}.
Microsoft's OSSGadget generates lexical string permutations and verifies the existence of mutated package names on SPRs to detect potential package confusion attacks~\cite{MicrosoftOSSGadget}.

Except for Stacklok's, these defenses rely solely on the characters in the package
name, whether through edit-distance heuristics or semantic similarity.
We refer to these lexical or semantic name features collectively as
  ``\textit{string-only signals}''.
These defenses ignore auxiliary metadata such as
  popularity,
  maintainers,
  and READMEs.


\subsection{Gaps in Current Approaches} \label{subsec:knowledge-gaps}

Despite the proposed approaches for detecting package confusion attacks, both theoretical and practical gaps remain. 

\subsubsection{Theoretical Gaps}
\label{sec:TheoreticalGaps}
Current detectors primarily depend on \textit{string-only signals}, disregarding contextual metadata.
This narrow view is likely to inflate false positive rates when legitimate projects share similar names, while allowing stealthy threats to go undetected when attackers use benign-looking package names combined with adversarially crafted metadata (\eg package descriptions, READMEs) to deceive engineers.
Furthermore, prior work's \textit{definition of package confusion} requires packages to contain active malware~\cite{neupane2023beyondTyposquatting}, failing to account for stealthy threats that initially appear benign but may become malicious later. This oversight highlights a gap in understanding the full threat landscape.

Additionally, detailed empirical treatments have so far focused exclusively on package confusion attacks in NPM, PyPI, and Ruby\-Gems \cite{taylor2020spellboundDefendingAgainstPackageTyposquatting,neupane2023beyondTyposquatting,vu2020typosquatting}, omitting package confusion attacks in SPRs with \textit{hierarchical naming structures} such as Maven, Golang, and Hugging Face.
This limited scope means that existing detection algorithms and similarity metrics may not generalize to SPRs with hierarchical names, which offer attackers additional attack surfaces through namespace manipulation and compound squatting techniques.

\subsubsection{Practical Gaps}
\label{sec:PracticalGaps}

The primary concern, as articulated by Ohm \etal~\cite{ohm2023SoKPracticalDetectionofSSCAttacks}, is that current package confusion detectors have \textit{high false-positive rates}.
Prior studies have shown that engineers consider monitoring systems with excessive false positives to be less useful than conservative ones~\cite{erlenhovEmpiricalStudyOfBots2020}.
In the context of package confusion detection, false positives
  cause several practical issues:
    they lead to alert fatigue among security analysts,
    reduce the likelihood that real threats will be investigated;
    erode trust in the detection system, causing organizations to disable or ignore alerts;
    and 
    harm legitimate package developers' reputations~\cite{erlenhov2020empirical}.

The engineering community acknowledges the severity of false positives. For example, developers of Microsoft's OSSGadget typosquatting detection tool explicitly seek ways to reduce false positives~\cite{2025improvements}.
Additionally, while industry reports mention package confusion defenses being used in production (\eg NPM and PyPI's lexical distance checks, Stacklok's risk scoring), these reports \textit{lack concrete deployment details} such as false positive rates, detection accuracy, number of threats caught, or operational challenges.
Without concrete performance metrics and deployment experiences, it remains unclear how effective these systems are in practice.

\section{Problem Statement}
\label{sec:ProblemStat}

In this section, we present an improved definition of package confusion threats (\cref{sec:ProblemStatement-Definition}), outline our threat model (\cref{sec:SystemandThreatModel}), and specify the system requirements necessary to effectively detect package confusion attacks (\cref{sec:ProblemState-SysReq}).

\subsection{Refined Definition of Package Confusion}
\label{sec:ProblemStatement-Definition}
We refine the definition of package confusion threat by emphasizing the manifestation of malicious intent.
Our refined definition is supported by our analysis of package confusion attacks (\cref{sec:PracticeAnalysis}).

\begin{tcolorbox}[title=Definition: Package Confusion Threat,boxsep=2pt, left=2pt, right=2pt, top=2pt, bottom=2pt]
\textbf{A package confusion threat} is a software package whose name mimics a trusted resource, with the intent of deceiving developers into installing code that is \ul{actively} malicious or \ul{may become so (stealthy)}.\footnotemark
\end{tcolorbox}

\footnotetext{For comparison, Neupane \etal wrote: \textit{``Package confusion attack: a malicious package is created that is designed to be confused with a legitimate target package and downloaded by mistake''}~\cite{neupane2023beyondTyposquatting}.
This requires the package to be actively malicious.} 

\noindent
As depicted in~\cref{fig:ThreatModel}, the mimicry may be \textit{\ul{lexical}} (\ie typosquatting), \textit{\ul{syntactic}}, or \textit{\ul{semantic}}, but the intent is to confuse an engineer.

By framing our definition around package confusion \textit{threats} rather than just \textit{attacks}, we admit the possibility that packages may initially appear benign but harbor malicious intent. 
In contrast, prior work relies on malware detectors and only finds actively malicious packages~\cite{taylor2020spellboundDefendingAgainstPackageTyposquatting,neupane2023beyondTyposquatting}.
Expanding the package confusion detection scope from active attacks to stealthy threats enlarges the candidate set of similar package pairs and, if done naively, can inflate false positives. 
We mitigate this challenge through our empirical analysis of false positive patterns in existing package confusion datasets (\cref{sec:ProblemState-CaseStudy2-FP}), which reveals distinguishing signals between genuinely benign packages and stealthy threats awaiting activation.






\subsection{Threat Model}
\label{sec:SystemandThreatModel}

We define the scope and boundaries of our threat model to clarify which package confusion attacks are addressed.
Our model encompasses both the context in which these attacks occur and the specific threats we aim to detect.

\textbf{Context Model:}
We focus on the SPRs mentioned in~\cref{sec:background-SPR-PkgNaming}, including both traditional SPRs hosting software packages and emerging platforms for pre-trained AI models.


\textbf{Threat Model:}
Our threat model distinguishes between package confusion attacks we address and those outside our scope:

\begin{itemize}[leftmargin=19pt]
    \item \textit{In-scope:} We consider attacks where adversaries publish packages with names deceptively similar to legitimate packages within SPRs.
    Consistent with our refined definition, attackers may initially upload benign content (stealthy threats) and later inject malicious code (active threats).
    \item \textit{Out-of-scope:} We exclude attacks that either use non-confusing package names or compromise existing legitimate packages through other means. 
    Such attacks require different mitigation strategies~\cite{okafor_sok_2022}.
\end{itemize}

\noindent
This threat model is stronger than those of existing package confusion detection approaches~\cite{vu2020typosquatting, taylor2020spellboundDefendingAgainstPackageTyposquatting, neupane2023beyondTyposquatting} because we account for stealthy attacks that begin with non-malicious package content.

\subsection{Requirements}
\label{sec:ProblemState-SysReq}
A system for detecting package confusion threats must meet both security and operational goals.
Working with our industry partner, we identified four requirements:

\noindent\textbf{\ul{Req$_{1}$}: High Precision and Recall.}
The system must balance
  precision (low false positive rate) against recall (low false negative rate).
False negatives leave deceptive packages unflagged, but false positives lead to alert fatigue~\cite{ahrq_alert_fatigue}.

\noindent\textbf{\ul{Req$_{2}$}: Efficient and Timely Detection.}
The system must scale to large SPRs while making low-latency checks for real-time feedback.

\noindent\textbf{\ul{Req$_{3}$}: Compatibility Across Ecosystems.}
The system must support many SPRs with their diverse naming structures. 


\noindent\textbf{\ul{Req$_{4}$}: High Recall of Stealth Package Confusion Attacks.}
The system must identify both active and stealthy package confusions.



\section{Analysis of Confusion Attacks}
\label{sec:EmpiricalAnalysis}

To address the high false positive rates of previous package confusion detection systems, Ohm \etal recently called for a deeper analysis of the available evidence~\cite{ohm2023SoKPracticalDetectionofSSCAttacks}.
We respond to their call in two ways:
  describing attackers' practices (\cref{sec:PracticeAnalysis}),
  and
  reporting metadata features that distinguish true from false positives (\cref{sec:ProblemState-CaseStudy2-FP}).

\subsection{Attackers' Practices}
\label{sec:PracticeAnalysis}

We examined first the role of stealth in confusion attacks, and then the attack variations in SPRs beyond NPM, PyPI, and RubyGems.
Our findings informed \ul{\textbf{Req$_{1,3,4}$}}.



\label{sec:ProblemState-CaseStudy1-TP}


\subsubsection{Stealthy Confusion Attacks}
We analyzed 240
confirmed NPM package confusion ``true positives'' (\ie confusing packages that include malware), based on data provided by Neupane \etal~\cite{neupane2023beyondTyposquatting}. We analyzed the number of days these malicious packages were available before malware was injected.
We estimated the injection time based on the last version updated before removal by the SPR.

\begin{figure}[ht]
    \centering
    \includegraphics[width=0.95\linewidth, trim=25 17 25 40, clip]{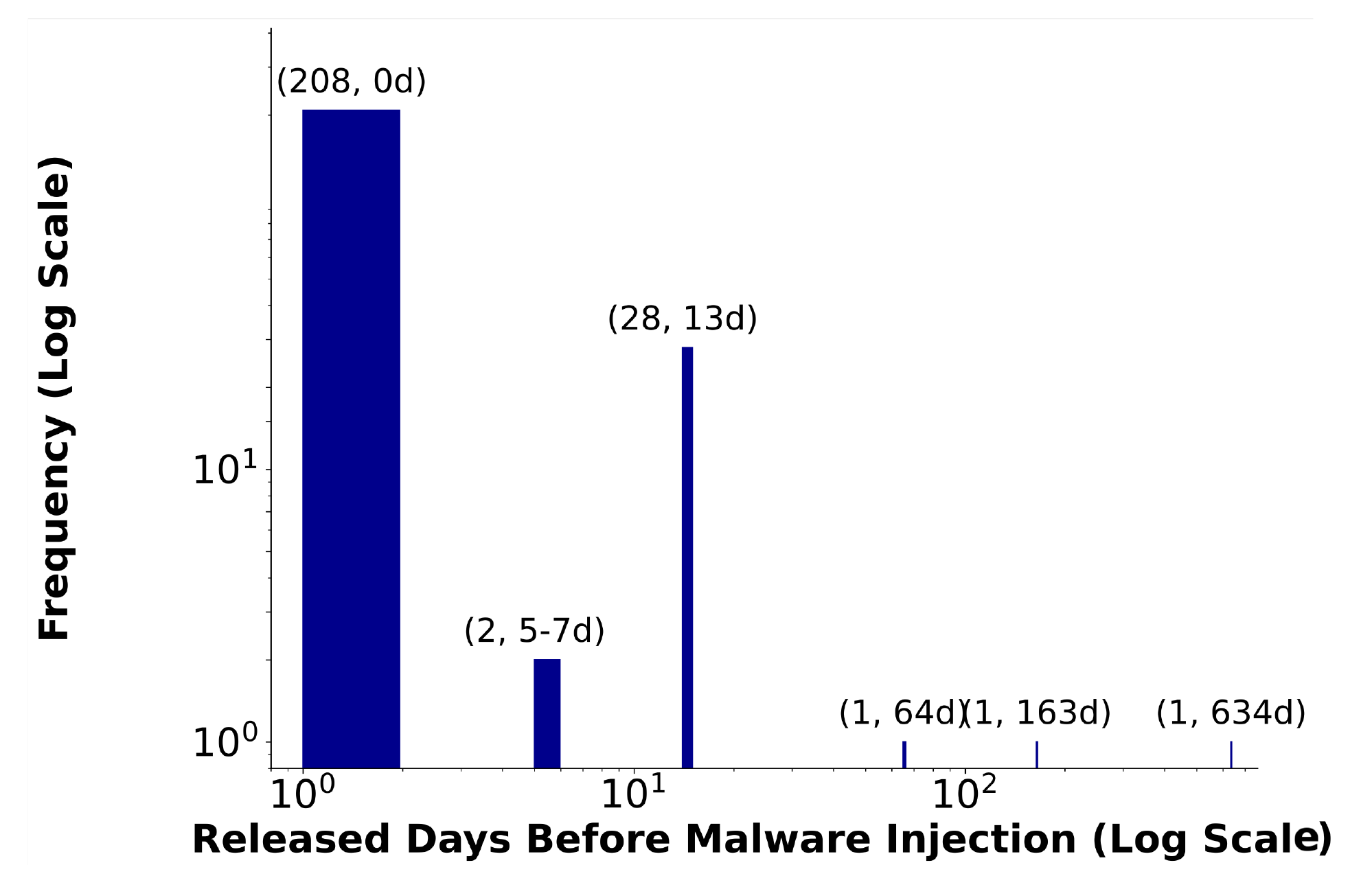}
    \caption{
    Distribution of time until malware, in packages with confusing names.
    13.3\% (32/240) of attacks occur $\geq$5 days after release.
    }
    \label{fig:TPAnalysis}
\end{figure}

\Cref{fig:TPAnalysis} illustrates our findings on the release time before malware injection.
Most package confusion attacks injected malware on the first day of release (\ie in the initial version).
However, in 13.3\% of the packages, attackers undertook stealth attacks, deploying malware after days or weeks.
Attackers may use this strategy to grow a user base before exploitation. 
Prior work on package confusion detection did not account for this behavior. 

\subsubsection{Extending the Taxonomy to New SPRs}
Prior work characterized malicious naming strategies in only a small set of SPRs, leaving gaps for ecosystems with hierarchical namespaces and platforms for hosting pre-trained AI models.
In collaboration with our industry partner's security analysts, who routinely triage package confusion incidents across all SPRs in \cref{tab:registry-overview}, we reviewed Neupane \etals taxonomy against their field observations.
Through this dialogue, we identified three additional mechanisms for package confusion attacks.
Definitions follow; see~\cref{tab:typosquat_taxonomy} for examples.

\begin{itemize}[leftmargin=*, itemsep=0.1ex]
  \item \textbf{Impersonation Squatting.}
    In hierarchical SPRs, attackers impersonate legitimate maintainers or organizations by registering a similar \texttt{author} or \texttt{groupId}.
  \item \textbf{Compound Squatting.}
    Making multiple coordinated edits to a hierarchical name, \eg altering both scope and delimiters at once.
    We consider this distinct because the compounded changes complicate conventional lexical comparisons.
  \item \textbf{Domain Confusion.}
    In SPRs where package names include URLs (\eg golang), attackers may register domains resembling those of trusted mirrors or proxies.
\end{itemize}

\subsection{Metadata Features of Package Confusion}
\label{sec:ProblemState-CaseStudy2-FP}




As discussed in~\cref{sec:background-Defenses}, almost all package confusion detectors make use only of the names of packages.
While this allows those systems to avoid costly content analysis, it leads to high false positive rates.
We propose to integrate metadata to distinguish true from false positives.
Although metadata signals can be circumvented, our approach aims to raise the attacker's cost, reflecting the typical cat-and-mouse dynamics in open package ecosystems~\cite{Ladisa2023TaxonomyofAttacksonOSSSC}.



{
\renewcommand{\arraystretch}{0.3}
\begin{table*}[ht]
\centering
\scriptsize
\setlength{\tabcolsep}{4pt}
\caption{
Overview of the 15 metadata-based verification rules.
Each rule lists its goal, the concrete implementation \& \textit{metadata} used (with inspected metadata fields \textit{italicized}), and the mapped LLM directive.  
In the \textbf{LLM Directive} column, ``---''  denotes a rule implemented purely with metadata checks and no LLM prompt. 
The final four rules (R12--R15)
were added during deployment to reduce false positives. 
}
\vspace{-0.10cm}
\label{tab:metadata_rules}
\scriptsize
\resizebox{\textwidth}{!}{%
\begin{tabular}{p{0.2\textwidth} p{0.34\textwidth} p{0.34\textwidth} p{0.11\textwidth}}
\toprule
\textbf{Rule} & \textbf{Description} & \textbf{Implementation \& \textit{metadata} used} & \textbf{LLM~Directive} \\
\midrule
\textbf{R1: Obviously Unconfusing}
& Determine if a package name conveys an intentional, brand-specific identity (e.g., \texttt{catplotlib}), clearly differentiating it from deceptive imitations.
& Cross-reference flagged \textit{package names} with known legitimate projects and use LLM analysis to assess whether the naming convention is deliberate and unambiguous for developers.
& \texttt{obvious\_not\_typosquat}\\
\\
\textbf{R2: Distinct Purpose}
& Distinguish packages with different functionalities, even if names are superficially similar (\eg \texttt{lodash-utils} \vs \texttt{lodash}).
& Pass both \textit{package descriptions} to the LLM; if the model reasons they serve genuinely different use cases or unique functionality, mark as distinct purpose, otherwise flag as possible impersonation. & \texttt{has\_distinct\_purpose} \\
\\
\textbf{R3: Fork Package}
& Detect benign forks sharing near-identical code or metadata with a popular package.
& Compare \textit{README files}, \textit{version histories}, and \textit{file structures} for high overlap. Similarities without malicious edits suggest harmless forks. & \texttt{is\_fork} \\
\\
\textbf{R4: Active Development/Maintained}
& Determine if packages are frequently updated or actively maintained by multiple contributors, which are less likely to have malicious intent.
& Retrieve \textit{last-update timestamp}, \textit{commit history}, and \textit{version count}. Classify packages with recent updates (\eg within 30 days) or more than five versions as legitimate. & --- \\
\\
\textbf{R5: Comprehensive Metadata}
& Packages missing critical metadata elements are suspicious. 
& Check for \textit{licenses}, \textit{maintainer information}, and \textit{repository links}.
& \texttt{no\_readme}
\\
\\
\textbf{R6: Overlapped Maintainers}
& Distinguish legitimate extensions or rebrands by verifying if the flagged package shares maintainers with the legitimate one.
& Match \textit{maintainer identifiers} (\eg email) between flagged and legitimate packages. Overlapping maintainers suggest legitimate intent. & \texttt{is\_known\_maintainer} \\
\\
\textbf{R7: Adversarial Package Name}
& Filter out name pairs with significant length differences, as these often indicate unrelated projects rather than covert mimicry.
& Compare flagged and legitimate \textit{package name lengths}. A difference exceeding 30\% indicates likely unrelated naming. & \texttt{is\_adversarial\_name}\\
\\
\textbf{R8: Well-known Maintainers}
& Trust packages maintained by reputable and recognized authors/organizations.
& Leverage knowledge in LLM training data to identify reputable \textit{maintainers} in the community.
& \texttt{is\_known\_maintainer} \\
\\
\textbf{R9: Clear Description}
& Legitimate packages document purpose and usage.
& Analyze repository metadata to ensure there are clear  \textit{package description} and \textit{README}; packages lacking such documentation should be flagged for further review. & \texttt{no\_readme} \\
\\
\textbf{R10: Has Malicious Intent}
& Identify packages that deliberately mimic legitimate ones through near-identical descriptions, or have obviously suspicious content.
& Use LLM to analyze the \textit{package name} and \textit{package description} to triage whether there is obvious malicious intent in the package. & \texttt{has\_suspicious\_intent} \\
\\
\textbf{R11: Experiment/Test Package}
& Detect packages that are used for test or experiment purposes only.
& Use LLM to analyze \textit{package description} and determine whether the package is for experimental use. & \texttt{is\_test} \\
\\
\midrule
\textbf{R12: Package Relocation}
& Account for legitimate package relocations, common in hierarchical registries like Maven and Golang.
& Parse \textit{project configuration files} (\eg look for \texttt{<relocation>} tag inside \texttt{pom.xml}). Identify and ignore renamed or migrated projects. & \texttt{is\_relocated\_package} \\
\\
\textbf{R13: Organization Allowed List}
& Prevent false positives by excluding packages published by trusted or verified organizations.
& Maintain a \textit{approved organizations list}. If a flagged package is published under an allowed organization (\eg \texttt{@oxc-parser/binding-darwin-arm64}), it should be considered legitimate, comparing to \texttt{binding-darwin-arm64}. & --- \\
\\
\textbf{R14: Domain Proxy/Mirror}
& Account for legitimate proxies or mirrors in the ecosystem, common in hierarchical registries like Golang.
& Maintain a \textit{recognized domains list} that serve as proxies or mirrors. If a given package is published under a valid domain (\eg \texttt{gopkg.in/go-git/go-git}), consider it legitimate when compared to its primary source (\texttt{github.com/go-git/go-git}). & --- \\
\\
\textbf{R15: Registered Prefix}
& Leverage NuGet’s prefix reservation: legitimate publishers can reserve name prefixes (\eg \texttt{Newtonsoft.Json.*}). 
& Nuget offers a \texttt{verified field} to indicate if a package name has a registered prefix. A verified package is considered legitimate.
& --- \\
\\
\bottomrule
\end{tabular}
}
\end{table*}
}

\subsubsection{Method}
We randomly sampled \SampledTypomindFPNum packages from the original \TypomindFPNum false positives reported by Neupane \etal~\cite{neupane2023beyondTyposquatting}, which have suspicious package names but are not flagged as malware in their respective security analyses.
This sample size gives a confidence level of \FPPrelimDataConfidenceLevel with a margin of error of \FPPrelimDataMarginError on the distribution. 

To identify possible metadata signals in the existing package confusion data,
we began by having two researchers 
analyze 200 of these packages.
They analyzed each package’s metadata (\eg READMEs, maintainers, and version timelines).
Each researcher independently proposed possible features based on this analysis (codebook) and then they discussed this codebook together to reach agreement on terms and definitions.
To test validity, they then independently applied this codebook to the 200 packages and measured agreement using Cohen's Kappa~\cite{cohen1960coefficient}.
The initial inter-rater reliability was \InterRaterAgreement (``substantial''~\cite{mchugh2012interrater}).
The researchers then convened to resolve discrepancies and refine their analysis. 
Through discussion, they reached a consensus on measurable attributes that could indicate malicious intent or the possibility of a stealthy attack.

Based on the high agreement in this process, one of these researchers analyzed the remaining 426 packages.

\subsubsection{Results}
Of the 626 packages sampled, 601 were still accessible.
Among these, we identified
  \FPBenignPkgs benign packages (\FPBenignPkgsPercent)
  and
  what we believe are \FPStealthyPkgs stealthy attacks (\FPStealthyPkgsPercent).
The progression of our results and the manually labeled dataset are available in our artifact (\cref{sec:OpenScience}).
From this analysis, we distilled 11 package metadata-based rules that differentiate genuine confusion attacks from benign look‑alikes.
These attributes include factors such as a distinct purpose, adversarial package naming, and the comprehensiveness of available metadata.
See~\cref{tab:metadata_rules} for \tool's rules.

\section{\tool Design and Implementation}
\label{sec:SystemDesign}

We introduce \tool, our detector for package confusion threats and attacks. 
\tool is designed to meet the system requirements defined in~\cref{sec:ProblemState-SysReq}. 
Leveraging the insights from our empirical study (\cref{sec:EmpiricalAnalysis}), \tool integrates
  syntactic and semantic package name analysis,
  hierarchical naming checks,
  and
  verifications based on package metadata to enhance threat detection.

\begin{figure*}[t]
    \centering
    \includegraphics[width=0.92\linewidth,
    trim=0 24.5 0 18,
    clip
    ]{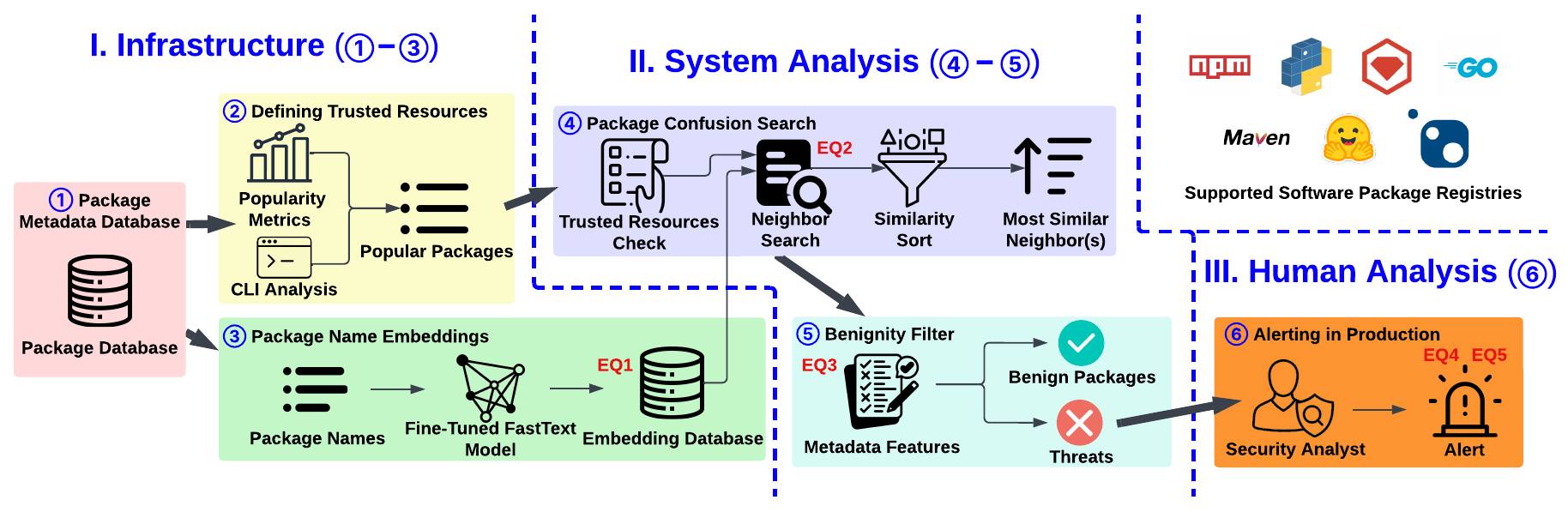}
    \vspace{-0.20cm}
    \caption{
    High-level architecture of \tool.
    There are three components divided into six parts.
    \textit{Component I -- Infrastructure}
      (\textcolor{blue}{\textcircled{\small{1}}}--\textcolor{blue}{\textcircled{\small{3}}})
      builds the package-metadata database (\cref{sec:SystemDesign-Step1}), identifies trusted resources (\cref{sec:SystemDesign-Step2}), and constructs the package name embedding model (\cref{sec:SystemDesign-Step3}).
    \textit{Component II -- System Analysis}
      (\textcolor{blue}{\textcircled{\small{4}}}--\textcolor{blue}{\textcircled{\small{5}}})
      performs the package-confusion search (\cref{sec:SystemDesign-Step4}) and benignity filtering (\cref{sec:SystemDesign-Step5}); these two parts are the main novelties of \tool.
    Components 1 and 2 together constitute the fully automated \tool pipeline.
    \textit{Component III -- Human Analysis} (\textcolor{blue}{\textcircled{\small{6}}}) routes possible package confusion attacks to security analysts for investigation (\cref{sec:SystemDesign-Step6}), whose decisions feed back into prompt, parameter, and threshold refinement to improve Components I and II.
    Red labels indicate where each evaluation question (EQ1--EQ5) is assessed.
    }
    \label{fig:pipeline}
\vspace{-0.25cm}
\end{figure*}

\cref{fig:pipeline} shows the components and parts of \tool: 

\begin{enumerate}[%
    label=\bfseries Component~\Roman*.,%
    align=left
]
  \item \textbf{Infrastructure:}
    \begin{enumerate}[label={},leftmargin=0pt,align=left]
      \item[{\StepRef{1}}] 
        A package metadata database, ensuring real-time awareness of new and evolving packages (\ul{\textbf{Req$_4$}}).
      \item[{\StepRef{2}}] 
        Popularity metrics to protect high-value targets (\ul{\textbf{Req$_{2,3}$}}).
      \item[{\StepRef{3}}] 
        A fine-tuned embedding model and database to capture domain-specific semantic name similarities (\ul{\textbf{Req$_{1,3}$}}).
    \end{enumerate}

  \item \textbf{System Analysis:}
    \begin{enumerate}[label={},leftmargin=0pt,align=left]
      \item[{\StepRef{4}}] 
        Syntactic and semantic confusion checks (\ul{\textbf{Req$_2$}}).
      \item[{\StepRef{5}}] 
        Filtering out benign packages using metadata (\ul{\textbf{Req$_{1,4}$}}).
    \end{enumerate}

  \item \textbf{Human Analysis:}
    \begin{enumerate}[label={},leftmargin=0pt,align=left]
      \item[{\StepRef{6}}] 
        Forwarding possible threats to security analysts (\ul{\textbf{Req$_{1,4}$}}).
    \end{enumerate}
\end{enumerate}

\vspace{-0.10cm}
\subsection{Part \textcolor{blue}{\textcircled{\small{1}}}: Package Metadata Database}
\label{sec:SystemDesign-Step1}

\subsubsection{Rationale}
Package confusion attacks are initiated regularly.
Up-to-date metadata facilitates threat detection.

\subsubsection{Approach}
\tool relies on a database with package names, version histories, commit logs, license info, maintainer records, and other publicly accessible metadata, all updated on a weekly basis.
Regular updates reduce concerns about stale data and ensure our results are up to date. 
We specifically use our industry partner's private database, which consolidates metadata from NPM, PyPI, RubyGems, Maven, Golang, Hugging Face, and NuGet.

\vspace{-0.10cm}
\subsection{Part \textcolor{blue}{\textcircled{\small{2}}}: Defining Trusted Resources}
\label{sec:SystemDesign-Step2}

\subsubsection{Rationale}
Confusion attacks mimic trusted resources.
If an attacker chooses a name that mimics an untrusted package, little harm can result.
Untrusted packages can be compared against trusted ones to focus on the threats of highest potential impact.


\subsubsection{Approach}
Like prior works~\cite{taylor2020spellboundDefendingAgainstPackageTyposquatting,neupane2023beyondTyposquatting}, we operationalize trust in terms of popularity.
Software engineers commonly make use of popularity signals as a proxy for trustworthiness~\cite{WhatsinaGithubStar}, and thus attackers choose names similar to popular packages.
Our specific popularity measure depends on the SPR.
Many SPRs --- in our case, NPM, PyPI, RubyGems, Hugging Face, and NuGet -- offer weekly or monthly download counts.
Our other two supported SPRs, Maven and Golang, do not publish download metrics.
For these, we make use of the \texttt{ecosyste.ms} database, which estimates popularity using indicators like stargazers, forks, and dependent repositories~\cite{ecosystemsWeb}.
The thresholds were set based on production experience and costs.
We initially chose
    5,000 weekly downloads for NPM, PyPI, RubyGems, and NuGet;
    1,000 for Hugging Face (normalized for its lower download rate~\cite{Jiang2022PTMSupplyChain,Jiang2022PTMReuse}); 
    and
    an \texttt{avg\_ranking} score of 10 for Maven and Golang.
We then adjusted the Golang thresholds to 4 to balance production latency.


Adversaries may inflate the popularity statistics of their packages~\cite{he2024FakeStars}, which could cause a malicious (or decoy) package to become trusted.
If our partner's security analysts flag any package as suspicious (\cref{sec:SystemDesign-Step6}), it is removed from the list of trusted packages.

\subsection{Part \textcolor{blue}{\textcircled{\small{3}}}: Package Name Embeddings}
\label{sec:SystemDesign-Step3}

\subsubsection{Rationale}
Detecting maliciously similar names requires know\-led\-ge of lexical, syntactic, and semantic variations.
Simple Levenshtein edit distance methods omit domain-specific semantic nuances (\eg \texttt{meta-llama} \vs \texttt{facebook-llama} on Hugging Face), while generic embedding models can introduce inaccuracies that
result in higher false positive or false negative rates.
Robust, fine-tuned embeddings address these shortcomings by providing enhanced semantic sensitivity, reducing erroneous alerts.
Prior work applies embedding for word-wise semantic similarity checks~\cite{neupane2023beyondTyposquatting}.
We also use embedding for the whole name to support efficient neighbor search (\cref{sec:SystemDesign-Step4}).
A generic embedding approach also supports various package naming conventions, \eg differences by SPR.

\subsubsection{Approach}
An \emph{embedding model fine-tuned on real package names} enhances semantic sensitivity, enabling the detection of adversarial or suspicious names. This capability is integral to assessing the risk associated with a package and serves as a cornerstone of our \textit{intent-centric} approach.
Following \cite{neupane2023beyondTyposquatting}, we build upon FastText~\cite{bojanowski2017FasttextEmbedding}, starting with the pre-trained model pre-trained on \texttt{cc.en.300.bin}, and \emph{fine-tune} it using all $\sim$9.1 million package names in our package metadata database as of Nov.~2024 (\S\ref{sec:SystemDesign-Step1}).
By fine-tuning, we expect to better capture domain-specific semantics.

We then use the fine-tuned embedding model to create an embedding database.
Given a package name, we remove its delimiters and compute the embedding of the concatenated string. 
For hierarchical names from
  Golang (\texttt{domain/author/repository}),
  Maven (\texttt{groupId:artifactId}),
  and
  Hugging Face (\texttt{author/model}), we split names to create distinct author and package identifier embeddings. 
This approach reduces false positives wherein similar packages from the same organization or author are incorrectly classified as confusing (\cref{sec:SystemDesign-Step4}).

The resulting embedding database occupies 24~GB.
We use the vector format provided by \texttt{pgvector} due to its efficient vector operations for databases~\cite{pgvector}.
This setup facilitates rapid query-based lookups
 and
supports subsequent steps in package neighbor searching.
Our artifact has visualizations of the embedding vectors (\cref{sec:OpenScience}).

\subsection{Part \textcolor{blue}{\textcircled{\small{4}}}:
Package Confusion Search}
\label{sec:SystemDesign-Step4}

Having obtained the appropriate infrastructure,
we must now detect package confusion threats efficiently yet accurately.
In this part, we focus on efficiently flagging possible package confusion threats. 

\subsubsection{Rationale}
Modern SPRs are enormous, so any detection scheme must be designed with scalability in mind. 
Given the set of all packages in an ecosystem, we follow prior work~\cite{taylor2020spellboundDefendingAgainstPackageTyposquatting, neupane2023beyondTyposquatting} by comparing the (relatively large) ``long tail'' of untrusted resources to the (relatively small) set of trusted resources.

\subsubsection{Approach}
To support the threshold choices and parameter settings used in \tool for detecting package confusion attacks, we studied the naming characteristics in these attacks.
We analyzed the 1,239 legitimate and typosquatted package name pairs from NPM, PyPI, and RubyGems provided by Neupane \etal~\cite{neupane2023beyondTyposquatting}.
Our analysis examined two key aspects of legitimate and typosquatted package name similarities: lexical distance using Levenshtein distance and semantic similarity using cosine similarity.
For semantic similarity measurements, we used cosine similarities of package names computed from our fine-tuned FastText embeddings (\cref{sec:SystemDesign-Step3}).

We confirmed that real-world package confusion attacks involve small syntactic changes and high semantic similarity.
The median Levenshtein distance between typosquatted and legitimate package names is 1, while the mean distance is 2.54. Additionally, 77.7\% of all package pairs exhibit a distance of 2 or less.
Among the analyzed package pairs, the mean cosine similarity was 0.95, and the median attained the maximum possible value of 1.

\myparagraph{Trusted Resources Check}
We compare untrusted resources (\ie unpopular packages) against all packages on the popular list.
For trusted resources (\ie popular packages), we restrict comparisons to ``more'' trusted packages.
A package is considered more trusted than a given trusted package if it has a download rate at least 10 times higher or an \texttt{ecosyste.ms} score at least twice as high --- following the criteria previously used by our industry partner.



\myparagraph{Neighbor Search (Package Name Similarity Search)}
We perform a distance search between each resource and the set of trusted resources.
Syntactic similarity is measured using Levenshtein distance (threshold$=2$), seeded by our empirical study and refined during deployment. 
For semantic similarity, we apply a cosine similarity threshold of 0.93, likewise initialized from the empirical baseline and finalized through deployment feedback.
As a special case for hierarchical names, we tighten the similarity thresholds to 0.99 for package names and 0.9 for author names to better detect ``compound squatting'' attacks (cf.~\cref{tab:detection_counts}); these tighter bounds reflect security analysts' observations during our field tests.

To efficiently measure semantic similarity, we leverage the \textit{Approximate Nearest Neighbors} method with the \textit{HNSW} index~\cite{malkov2018HNSW} in PostgreSQL. We opted for \textit{HNSW} over \textit{IVFFlat} based on benchmarking results that show faster search speeds and minimal vector perturbation~\cite{tembo_pgvector_2024}. By partitioning the embedding space into multiple clusters, the \textit{HNSW} index limits distance computations to a smaller subset of candidate packages, rather than exhaustively comparing all pairs (\ul{\textbf{Req$_{2}$}}).

\myparagraph{Similarity Sort and Most Similar Neighbor(s)}
Inspired by Neupane \etal~\cite{neupane2023beyondTyposquatting}, we define a similarity function that ranks neighbors based on Levenshtein distance, n-gram similarity, phonetics, substring matching, and fuzzy ratio.
We treat the nearest neighbors as the most likely targets.
The suspicious package and these nearest neighbors are propagated to the benignity check (\cref{sec:SystemDesign-Step5}).

In production, similarity sorting often returns multiple packages with near-perfect scores.
\cref{sec:EQ5} shows that adding more neighbors substantially increases latency and the cost of the benignity check, while delivering minimal accuracy gains.
Based on our industry partner's feedback, we restrict retrieval to the top-2 neighbors to optimize the trade-off between accuracy and speed.

\vspace{-0.1cm}
\subsection{Part \textcolor{blue}{\textcircled{\small{5}}}: Benignity Filter}
\label{sec:SystemDesign-Step5}


\subsubsection{Rationale}
Prior work used purely string-oriented methods, which often misclassify harmless or beneficial packages as confusion attacks.
To mitigate false positives, we developed a metadata-driven benignity check inspired by a recent metadata-aware malware detector~\cite{sun2024MetadataFeaturesforMalwareDetection}.
We seek to filter out legitimate packages based on explainable heuristics, avoiding unnecessary alerts. 


\subsubsection{Approach}
Our analysis of false-positive data (\cref{sec:ProblemState-CaseStudy2-FP}) showed many reasons why legitimate packages may have names resembling other trusted resources.
Initially, we iteratively developed 11 rules that utilize metadata features
(\eg version history, maintainers)
to distinguish package confusion attacks from benign packages.
We apply these rules to assess the benignity of package neighbors (\cref{sec:SystemDesign-Step4}) in order to reduce the false positive rate.
We added several additional rules during deployment.
\cref{tab:metadata_rules} summarizes the goals and implementation for each rule.
Once our name-based detector flags a package 
  we retrieve its metadata and apply benignity check rules.
A risk score is calculated using a simple weighted sum, which facilitated analyst feedback compared to a learned approach. 

\ifARXIV
\begin{listing}[ht]
    \begin{lstlisting}
# ------------ system_prompt (excerpt) ------------
You are an AI assistant helping to identify false positives in potential typosquat packages.
Analyze the given package names and descriptions to determine if a package is a legitimate alternative or a malicious typosquat.
\textbf{DECISION FRAMEWORK}
\textbf{A. Package Name Analysis (is_adversarial_name):}
  - TRUE if: Names are confusingly similar.
  - FALSE if: Names are clearly distinct or branded
\textbf{B. Legitimate Package Determination (obvious_not_typosquat):}
  - TRUE if: High edit distance AND distinct purpose AND good docs
  - FALSE if: Low edit distance OR similar purpose OR poor docs
\textbf{C. Package Categorisation Flags:}
  - \textbf{is_fork:} TRUE if package metadata explicitly states it is a fork
  - \textbf{has_distinct_purpose:} TRUE if packages solve clearly different tasks
  - \textbf{is_test:} TRUE if package name/description explicitly mentions testing

# ------------ user_prompt (excerpt) --------------
\textbf{PACKAGE COMPARISON ANALYSIS}
\textbf{Target Package:} {typo_name}
\textbf{Target Package Metadata:} {typo_metadata}
\textbf{Legitimate Package Metadata:} {legit_name}
\textbf{Legitimate Package Metadata:} {legit_metadata}
\textbf{Ecosystem:} {registry}

Analyze the information above and answer the questions below using TRUE or FALSE.

1. \textbf{obvious_not_typosquat:} Is '{typo_name}' clearly NOT a typosquat?
2. \textbf{is_adversarial_name:} Can users confuse '{typo_name}' with '{legit_name}'?
3. \textbf{is_fork:} Is '{typo_name}' a legitimate fork or variant?
...
9. \textbf{is_relocated_package:} Is this a known package relocation?
    \end{lstlisting}
    \vspace{-0.25cm}
    \caption{Excerpts of the system and user prompts used in the benignity filter. Full prompts are in our artifact (\cref{sec:OpenScience}). We refined both prompts iteratively with feedback from security analysts, ensuring that the LLM’s analysis aligns with the rule set in~\cref{tab:metadata_rules}.}
    \label{lst:llm-prompt}
  \end{listing}
\else
\begin{listing}
\begin{minted}[fontsize=\scriptsize, breaklines,breaksymbolleft=,escapeinside=||,frame=lines]{markdown}
# ---------------- system_prompt (excerpt) ----------------
You are an AI assistant helping to identify false positives in potential typosquat packages.
Analyze the given package names and descriptions to determine if a package is a legitimate alternative or a malicious typosquat.
|\textbf{DECISION FRAMEWORK}|
|\textbf{A. Package Name Analysis (is\_adversarial\_name):}|
  - TRUE if: Names are confusingly similar.
  - FALSE if: Names are clearly distinct or branded
|\textbf{B. Legitimate Package Determination (obvious\_not\_typosquat):}|
  - TRUE if: High edit distance AND distinct purpose AND good docs
  - FALSE if: Low edit distance OR similar purpose OR poor docs
|\textbf{C. Package Categorisation Flags:}|
  - |\textbf{is\_fork:}| TRUE if package metadata explicitly states it is a fork
  - |\textbf{has\_distinct\_purpose:}| TRUE if packages solve clearly different tasks
  - |\textbf{is\_test:}| TRUE if package name/description explicitly mentions testing
# ---------------- user_prompt (excerpt) ------------------
|\textbf{PACKAGE COMPARISON ANALYSIS}|
|\textbf{Target Package:}| {typo_name}
|\textbf{Target Package Metadata:}| {typo_metadata}
|\textbf{Legitimate Package Metadata:}| {legit_name}
|\textbf{Legitimate Package Metadata:}| {legit_metadata}
|\textbf{Ecosystem:}| {registry}

Analyze the information above and answer the questions below using TRUE or FALSE.

1. |\textbf{obvious\_not\_typosquat:}| Is '{typo_name}' clearly NOT a typosquat?
2. |\textbf{is\_adversarial\_name:}| Can users confuse '{typo_name}' with '{legit_name}'?
3. |\textbf{is\_fork:}| Is '{typo_name}' a legitimate fork or variant?
...
9. |\textbf{is\_relocated\_package:}| Is this a known package relocation?
\end{minted}
\vspace{-0.25cm}
\caption{
Excerpts of the system and user prompts used in the benignity filter.
Full prompts are in our artifact (\cref{sec:OpenScience}).
We refined both prompts iteratively with feedback from security analysts, ensuring that the LLM's analysis aligns with the rule set in~\cref{tab:metadata_rules}.
}
\label{lst:llm-prompt}
\end{listing}
\fi

To realize this filter, we first distinguished the rules that can be assessed using Boolean algebra over package metadata (\eg R4), from those that require extensive analysis (\eg R2) or subjective reasoning (\eg R9).
The former we implemented with conventional software, and for the latter we used an LLM.
We tried two prompting approaches: 
  rule-based
  and 
  ``LLM-as-judge''~\cite{gu2025surveyllmasajudge}.
For rule-based, we provided detailed rules imitating security analysts' reasoning (\eg specific metadata they referenced).
This detected known attacks (accuracy of 0.82 on NeupaneDB),
 but
 overfit, raising false positives on new data (accuracy of 0.53 on \texttt{ConfuDB} as of March 2025).
We also tried the ``LLM-as-a-Judge'' pattern, casting the model as an evaluator guided by a structured decision framework but without excessive detail~\cite{gu2025surveyllmasajudge}.
This pattern performed better, with accuracy of 0.83 on NeupaneDB and 0.68 on the latest \texttt{ConfuDB} (\cref{tab:EQ5_Accuracy}). 

\Cref{lst:llm-prompt} shows the final system and user prompts. 
We used OpenAI's \texttt{o4-mini} in our implementation. 

\vspace{-0.1cm}
\subsection{Part \textcolor{blue}{\textcircled{\small{6}}}: Alerting in Production}
\label{sec:SystemDesign-Step6}


\myparagraph{Use in Production}
\tool is integrated into our industry partner's security scanning system for software packages.
They had previously relied primarily on content-based malware scanners, but these are too computationally costly to scale well. 
The partner's former approach for first-pass detection of confusion attacks was a lexical check (Levenshtein distance) which had a
  high false positive rate.
\tool replaced that approach. 
Analysts triage alerts from \tool and the other scanners to make a threat assessment.
If they deem a package to be malicious, that information is propagated to our partner's customers via security feeds.


\myparagraph{Experience-Driven Optimizations}
During deployment we made several optimizations. 
We share three examples.
\textit{First}, we enhanced our metadata checks by analyzing false positives and adding rules \texttt{R11}--\texttt{R15} to better distinguish benign packages (\cref{tab:metadata_rules}). 
\textit{Second}, we improved accuracy by splitting long identifiers,
  such as
  Golang’s domain-based names,
  into their constituent parts and computing similarity on each part. 
\textit{Third}, we found the LLM-as-Judge pattern was more effective than detailed prompts (\cref{sec:SystemDesign-Step5}).

\section{Evaluation}
\label{sec:Eval}

To evaluate \tool, we pose five Evaluation Questions (EQs) to assess its performance at the component level (\textbf{EQ1}–\textbf{EQ3}) and the system level (\textbf{EQ4}–\textbf{EQ5}).
At the component level, we measure the effectiveness of novel mechanisms introduced in \tool.
At the system level, we evaluate its integrated functionality and scalability compared to baseline tools (\cref{sec:baselines}), and ability to detect real-world package confusion threats.
Additionally, we compare our approach to SOTA methods to benchmark its effectiveness. An overview of the evaluation process is illustrated in \cref{fig:pipeline}.



\myparagraph{Component-level}
We assess how individual components contribute effectively to the overall system.

\begin{itemize}[leftmargin=*, itemsep=0.1ex]
    \item \textbf{EQ1: Performance of Embedding Model.}
    What is the accuracy and efficiency of our
    embedding model? (\cref{sec:SystemDesign-Step3})

    \item \textbf{EQ2: Neighbor Search Accuracy.}
    How effective is the neighbor search approach?
    (\cref{sec:SystemDesign-Step4})

    \item \textbf{EQ3: Benignity Check Accuracy.}
    How much does the benignity filter reduce false positive rates? (\cref{sec:SystemDesign-Step5})
\end{itemize}

\myparagraph{System-level}
We examine the performance of the full \tool system and compare to other approaches.

\begin{itemize}[leftmargin=*, itemsep=0.1ex]
    \item \textbf{EQ4: Discovery.}
    Can \tool identify previously unknown package confusion threats?
    \item \textbf{EQ5: Baseline Comparison.}
    How does \tool perform in terms of accuracy and latency compared to SOTA tools?



\end{itemize}

All experiments run on a server with 32 CPU cores (Intel Xeon CPU @ 2.80GHz) and 256 GB of RAM.
Notably, the training and fine-tuning of FastText models do not require GPUs.



\subsection{Baseline and Evaluation Datasets}
\label{sec:Eval-Setup}

\myparagraph{State-of-the-Art Baselines}
\label{sec:baselines}
We compare our system to the Levenshtein distance approach~\cite{vu2020typosquatting}, and
Typomind~\cite{neupane2023beyondTyposquatting}.
We consider OSSGadget~\cite{MicrosoftOSSGadget} out of scope because it only handles lexical confusions and its latency for long package names makes it unsuitable for production. These are all state-of-the-art open-source tools.

\myparagraph{Evaluation Datasets}
We evaluate using two datasets:

\begin{itemize}[leftmargin=1.5em, itemsep=0.1ex]
    \item \texttt{\textbf{NeupaneDB}}:
    \NumNeupaneDB packages from \cite{neupane2023beyondTyposquatting}, including
        \NumTypomindData confirmed real-world package confusion attacks,
        and
        \SampledAvailableTypomindFPNum manually labeled data we analyzed in \cref{sec:ProblemState-CaseStudy2-FP}.
\item  \texttt{\textbf{ConfuDB}}: \NumConfuDB packages triaged by security analysts, collected during the development and refinement of \tool (\cref{sec:SystemDesign-Step6}).
\end{itemize}

\subsection{EQ1: Performance of Embedding Model}
\label{sec:EQ1}

\ifNOBOXES
\begin{tcolorbox} [width=\linewidth, colback=yellow!30!white, top=1pt, bottom=1pt, left=2pt, right=2pt]
\textbf{EQ1 Summary:}
Our embedding model demonstrates superior effectiveness and efficiency, ensuring an acceptable overhead for SPR during the preparation of the embedding database.
\end{tcolorbox}
\fi

We measured effectiveness and efficiency of our embedding model.
\vspace{-0.15cm}

\subsubsection{Effectiveness of Embedding Model}
\myparagraph{Method}
We evaluate the effectiveness of embedding-based similarity detection by comparing three approaches:

\begin{enumerate}[leftmargin=*, itemsep=0.2ex]
    \item \textit{Levenshtein Distance}, calculating the number of single-character edits required to change one package name to another.
    \item \textit{Pre-trained FastText}~\cite{bojanowski2017FasttextEmbedding} (\texttt{cc.en.300.bin}),
    as used in the SOTA work Typomind to capture semantic relationships~\cite{neupane2023beyondTyposquatting}.
    \item \textit{Fine-tuned FastText} (\textbf{Ours}), trained with a corpus of package names to better capture domain-specific similarities.
\end{enumerate}

To systematically compare these methods, we construct a balanced test set consisting of both positive and negative pairs, with each category containing 1,239 data points~\cite{Bishop06PRML, duda2006pattern}.

\begin{itemize}[leftmargin=*, itemsep=0.2ex]
    \item \textit{Positive Pairs}: All 1,239 of the real package confusions from \texttt{NeupaneDB}, which were confirmed and removed from SPRs~\cite{neupane2023beyondTyposquatting}.

    \item \textit{Negative Pairs}: Created synthetically by randomly pairing unrelated package names from registries, ensuring low similarity scores across all tested methods.
    These pairs are guaranteed not to represent package confusion relationships.
\end{itemize}

For each approach, we applied a predefined similarity threshold to classify package pairs as potential package confusion attacks. 
The threshold was selected via a grid search
 over values from 0 to 1 in 0.01 increments
to optimize Precision and Recall, ensuring effective identification of true confusions.
Pairs with similarity scores above the threshold were classified as positive, while those below were classified as negative.
False positives were subsequently filtered using our false-positive verification process (\cref{sec:SystemDesign-Step5}).
We compare the Precision, Recall, and F1 scores for each approach. 

\myparagraph{Result}
\cref{tab:EQ1_effectiveness_updated} presents the comparison between our fine-tuned model and baseline approaches. The results show that our approach outperforms both the Levenshtein Distance and Pretrained models, achieving the highest F1 score for positive pairs (0.95) while maintaining strong performance on negative pairs (0.98).

\renewcommand*{\arraystretch}{0.5} \begin{table}[ht]
\small
\centering
\caption{
Similarity effectiveness of using distance and embedding. We apply a grid search which automatically determines the optimal threshold to maximize F1 scores for positive pairs while maintaining relatively high F1 scores for negative pairs.
}
\vspace{-0.1cm}
\label{tab:EQ1_effectiveness_updated}
\resizebox{0.95\linewidth}{!}
{
\small
\begin{tabular}{lccccccc}
\toprule \textbf{Model} & \multicolumn{3}{c}{\textbf{Positive Pairs}} & \multicolumn{3}{c}{\textbf{Negative Pairs}} & \textbf{Overall Score} \\
\cmidrule(lr){2-4} \cmidrule(lr){5-7} & P & R & F1 & P & R & F1 & \\
\midrule
Levenshtein Distance & 1.00 & 0.80 & 0.89 & 1.00 & 1.00 & 1.00 & 0.94 \\
Pretrained & 1.00 & 0.85 & 0.92 & 1.00 & 0.98 & 0.99 & 0.96 \\
Fine-Tuned (\textbf{Ours}) & 1.00 & \textbf{0.90} & \textbf{0.95} & 1.00 & 0.96 & 0.98 & \textbf{0.96} \\

\bottomrule
\end{tabular}
}
\end{table}

We performed a grid search and analyzed the ROC curve to assess how the similarity threshold affects accuracy and to plot the ROC curve of our fine-tuned model. This analysis identified 0.93 as the optimal threshold, which we adopted in production. With an AUC of 0.94, the model demonstrates a balanced trade-off between precision and recall for both positive and negative pairs.

\subsubsection{Efficiency of Embedding Model}

\myparagraph{Method}
To assess the efficiency of embedding database creation, we evaluated the cost of creating the embedding database (\cref{sec:SystemDesign-Step3}).
We measured the throughput, latency, and overall overhead associated.

\myparagraph{Result}
Creating the embedding database can be done in a reasonable amount of time, even for large SPRs.
Total database preparation cost varied from 28--650 seconds (RubyGems, NPM), largely driven by the size of the dataset in each ecosystem.  
Throughputs ranged from $\sim$4.9--13.6\ K pkgs/sec (Maven, PyPI), with per-package latencies consistently below 0.2 ms.
HNSW indexing adds $<$\ 8 seconds of upfront processing for each ecosystem.





\subsection{EQ2: Neighbor Search Accuracy}
\label{sec:EQ2}
\ifNOBOXES
\begin{tcolorbox} [width=\linewidth, colback=yellow!30!white, top=1pt, bottom=1pt, left=2pt, right=2pt]
\textbf{EQ2 Summary:}
Our embedding-based neighbor search algorithm, combined with the sorting algorithm, achieves accuracy comparable to prior work while capturing richer semantic information.
\end{tcolorbox}
\fi

\myparagraph{Method} We use a threshold of 0.93 obtained from EQ1. To evaluate neighbor search performance, we analyzed suspicious packages identified by \texttt{Typomind}.

\myparagraph{Result}
Our neighbor search algorithm accurately detected 99\% of the \NumTypomindData real-world package confusion attacks from \texttt{NeupaneDB}~\cite{neupane2023beyondTyposquatting}.
Moreover, our method effectively flagged impersonation squatting attacks targeting hierarchical package names, notably identifying reported cases on Hugging Face \cite{protectai_huggingface} and Golang \cite{socketMaliciousGoModule} that earlier methods overlooked. These findings confirm that our neighbor search algorithm achieves state-of-the-art performance.

\subsection{EQ3: Benignity Check Accuracy}
\label{sec:EQ3}

\ifNOBOXES
\begin{tcolorbox} [width=\linewidth, colback=yellow!30!white, top=1pt, bottom=1pt, left=2pt, right=2pt]
\textbf{EQ3 Summary:}
Our metadata verification process significantly lowers the false-positive rate from 75.4\% to 5\%.
\end{tcolorbox}
\fi

\myparagraph{Method}
Using the raw embedding output (\ie packages flagged for name-based suspicion), we apply the benignity check from~\cref{sec:SystemDesign-Step5}.

We first use cross-validation to learn the weights and measure false positive rates (FPR) using \texttt{NeupaneDB}.
We then computed SHAP values~\cite{lundberg2017unifiedSharpley} by learning the weighed sum through regression to quantify the importance of each metric defined in \cref{tab:typosquat_taxonomy}. The resulting SHAP plot provides clear insights into how these metrics contribute to our system's decision-making process.
Additionally, we use regression to learn the weights.

\begin{table*}[ht]
\centering
\caption{
Accuracy metrics for \tool and \texttt{Typomind}, detailing true/false positives and negatives, as well as recall, precision, F1 score, and accuracy for active, stealthy, and benign threats.
Although Typomind flags all packages as suspicious, our benignity filter substantially reduces false positives in production.
Due to availability of metadata, we evaluated \tool on \NumNeupaneDBConfuguardEvaluated packages from \texttt{NeupaneDB} and \NumConfuDBConfuguardEvaluated from \texttt{ConfuDB}.
We note that \tool exhibits lower benign classification accuracy in \texttt{ConfuDB} because it flags more complex cases (\eg compound squatting) as potential threats, a nuance that \texttt{Typomind} does not capture.
ConfuDB data was collected between January and June 2025.
}
\vspace{-0.1cm}
\label{tab:EQ5_Accuracy}
\resizebox{0.95\linewidth}{!}
{
\centering
\begin{tabular}{cccr|rrrrcccc|rrrrcccc}
\toprule
\textbf{Dataset} & \textbf{Total} & \textbf{Threat Type} & \textbf{Sub-total} 
  & \multicolumn{8}{c|}{\tool (\textbf{Ours})} 
  & \multicolumn{8}{c}{\texttt{Typomind}} \\
\cmidrule(lr){5-12} \cmidrule(lr){13-20}
 &  &  &  
  & \textbf{TP} & \textbf{FP} & \textbf{TN} & \textbf{FN} 
  & \textbf{Recall} & \textbf{Prec.} & \textbf{F1} & \textbf{Acc.} 
  & \textbf{TP} & \textbf{FP} & \textbf{TN} & \textbf{FN} 
  & \textbf{Recall} & \textbf{Prec.} & \textbf{F1} & \textbf{Acc.} \\
\midrule
& & Active   & 1,239 & 1,103 &   0 &   0 & 131 & 0.89 & 1.00 & 0.94 & 0.89 & 1,000 &   0 &   0 & 239 & 0.80 & 1.00 & 0.89 & 0.81 \\
NeupaneDB & \NumNeupaneDB & Stealthy &  121 &    68 &   0 &   0 &  53 & 0.56 & 1.00 & 0.72 & 0.56 &   121 &   0 &   0 &   0 & 1.00 & 1.00 & 1.00 & 1.00 \\
& & Benign   &  480 &     0 & 117 & 347 &   0 & 0.00 & 0.00 & 0.00 & 0.72 &     0 & 480 &   0 &   0 & 0.00 & 0.00 & 0.00 & 0.00 \\
\cmidrule(lr){3-4}
& & \textit{Overall} & \NumNeupaneDB 
  & 1,171 & 117 & 347 & 184 
  & \textbf{0.86} & \textbf{0.90} & \textbf{0.88} & \textbf{0.83}
  & 1,121 & 480 &   0 & 239 
  & 0.82 & 0.70 & 0.76 & 0.61 \\
\midrule
& & Active   & 133 & 61 & 0 & 0 & 72 & 0.53 & 1.00 & 0.70 & 0.53 & 71 & 0 & 0 & 62 & 0.53 & 1.00 & 0.72 & 0.56 \\
ConfuDB & \NumConfuDBConfuguardEvaluated & Stealthy & 409 & 238 & 0 & 0 & 171 & 0.74 & 1.00 & 0.85 & 0.73 & 219 & 0 & 0 & 190 & 0.64 & 1.00 & 0.78 & 0.64 \\
& & Benign & 1,492 & 0 & 414 & 1078 & 0 & 0.00 & 0.00 & 0.00 & 0.72 & 0 & 560 & 932 & 0 & 0.00 & 0.00 & 0.00 & 0.61 \\
\cmidrule(lr){3-4}
& & \textit{Overall} & \NumConfuDBConfuguardEvaluated & 299 & 414 & 1078 & 243 & \textbf{0.55} & \textbf{0.42} & \textbf{0.48} & \textbf{0.68} & 290 & 560 & 932 & 252 & 0.62 & 0.28 & 0.39 & 0.62 \\

\bottomrule
\end{tabular}
}
\end{table*}

\begin{figure}[ht]
    \centering
    \includegraphics[width=0.90\linewidth, trim=35 35 35 35, clip]{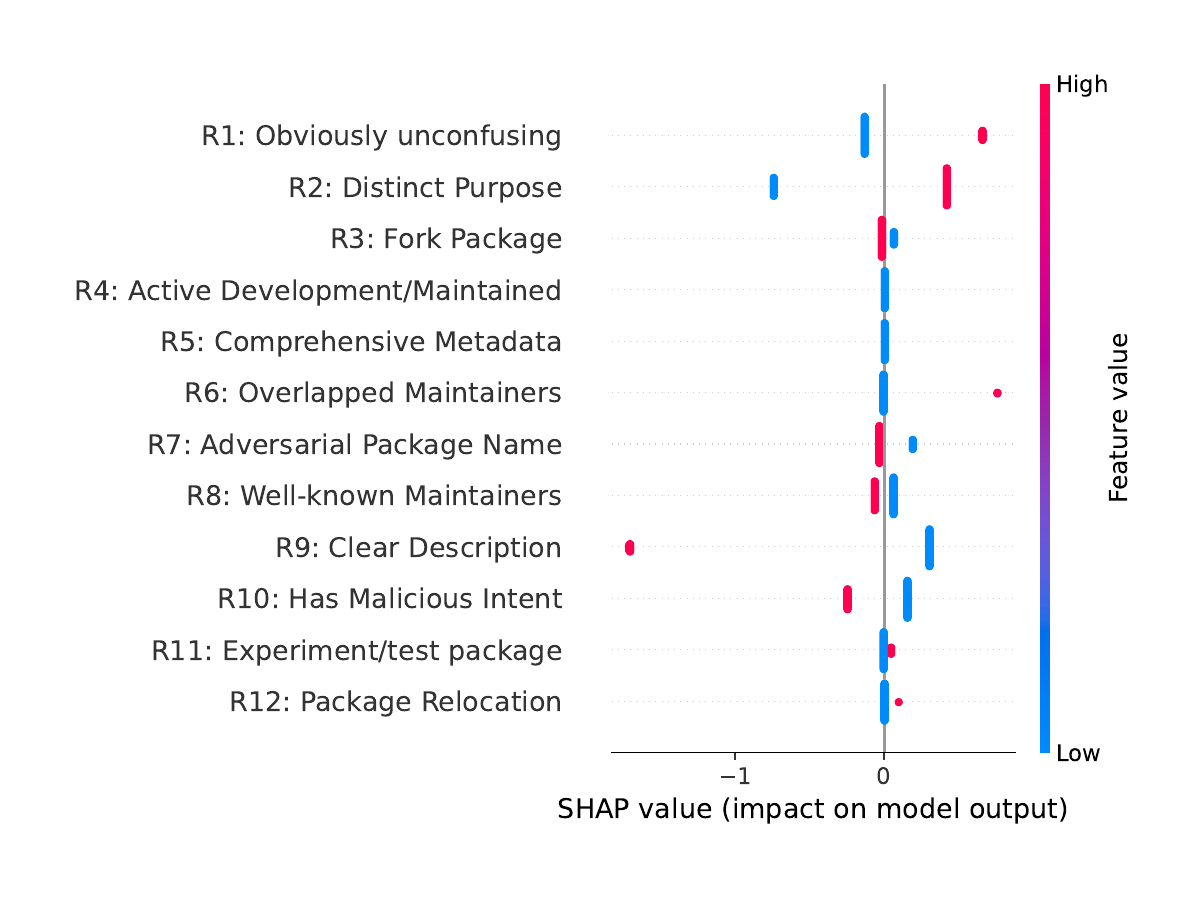}
    \vspace{-0.1cm}
    \caption{
    SHAP value plot for metadata verification over the learnable features from~\cref{tab:metadata_rules} (\ie excluding R13, R14, R15).
    The x-axis shows each feature's impact on model output, where positive values are towards benignity.
    Feature values are color-coded: red indicates high values with a strong, direct impact on the model output.
    }
    \label{fig:metadata-ablation}
\end{figure}

%
\myparagraph{Result}
%
\cref{fig:metadata-ablation} shows the SHAP value plot which indicates that our rules used in the benignity filter are useful. The plot highlights that distinct purpose (R2) and suspicious intent (R10) are the most influential features driving the model’s predictions, suggesting that packages showing suspicious or unusual behavior strongly push the model toward a malicious classification. Conversely, features such as known maintainers (R8) and fork packages (R3) generally pull the model’s output toward benign, indicating they lower the likelihood of a threat. Meanwhile, no clear description also shows a notable positive effect, aligning with the idea that missing documentation can be an indicator of malicious intent (R9).

We use cross-validation to learn parameters and estimate the trade-off between reducing false positives and increasing false negatives. Our benignity filter correctly flagged 347 out of 480 (72\%) false positives in the human labeled set. In the real-world attack set, it failed to classify 131 true positives (10\%) out of the 1,234 that had available metadata. 
A detailed investigation revealed two main causes:
  (1) missing metadata,
  and
  (2) attacks that had been removed, with the package now maintained by npm, which the filter considers a valid maintainer.




These results confirm that by including supplementary heuristics beyond raw name similarity,
false positive rates can be reduced while retaining high recall for genuinely malicious confusions.

\subsection{EQ4: Discovery of New Package Confusions}
\label{sec:EQ4}

\ifNOBOXES
\begin{tcolorbox} [width=\linewidth, colback=yellow!30!white, top=1pt, bottom=1pt, left=2pt, right=2pt]
\textbf{EQ4 Summary:}
Our system effectively detected stealthy typosquatting attacks with malicious intent and identified two new attack types—\textit{impersonation squatting} and \textit{compound squatting}—in a production environment. Moreover, it uncovered these threats across three ecosystems that had not been explored in prior research.
\end{tcolorbox}
\fi


\myparagraph{Method}
To evaluate the effectiveness of our package confusion detection system, we deployed \tool in a production environment for three month. During this period, flagged packages were analyzed using a commercial malware scanner and reviewed by threat analysts for detailed insights.

\myparagraph{Result}
%
\cref{tab:EQ5_Accuracy} summarizes the package confusion attacks identified by \tool during the January–June 2025 deployment period (\ie \texttt{ConfuDB}).
Within this database, 1,686 packages were labeled as false positives.\footnote{The high false positive rate in \texttt{ConfuDB} is primarily due to the difficulty of confirming stealthy confusion attacks, which were left unreviewed by analysts.}
Among the \NumReviewedThreats remaining suspicious packages, 119 (19\%) were identified as executable malware, 7 (1\%) launched HTTP-flood or DDoS attacks, 6 (1\%) loaded external scripts from unknown URLs or CDNs, and 1 (\textless 1\%) harvested analytics data and sent it to the attackers, and 497 (79\%) carried out stealthy confusion attacks without obvious malicious content.

\cref{tab:detection_counts} shows the different kinds of package confusion threats identified by \tool during its 6 months of production use.
\tool effectively detects advanced threats (\cref{sec:PracticeAnalysis}) --- including compound and impersonation squatting --- across platforms such as Maven, Golang, and Hugging Face.

We share four examples of the \NumReviewedThreats attacks discovered by \tool and confirmed by analysts.
\textbf{\textit{(1)}} \tool identified an \textit{impersonation squatting} attack on Maven: the malicious package \texttt{\seqsplit{io.github.leetcrunch:scribejava-core}} mimicked the legitimate \texttt{com.github.scribejava}. Although the package content appears identical at first glance, it contains code that injects a network call to steal user credentials.
\textbf{\textit{(2)}} \tool detected malware on a Golang package, where the username \texttt{boltdb-go} attempted to impersonate \texttt{boltdb}.
\tool also flagged suspicious Hugging Face packages, such as \texttt{\seqsplit{TheBlock/Mistral-7B-Instruct-v0.2-GGUF}} (with only 217 downloads), mimicking benign package, \texttt{\seqsplit{TheBloke/Mistral-7B-Instruct-v0.2-GGUF}}, which has 92K monthly downloads.
\textbf{\textit{(3)}} \tool also captured \textit{domain confusions}, such as Go packages from \texttt{github.phpd.cn} and \texttt{github.hscsec.cn} which look like proxies of official Golang packages but contain malware that either sets up a rogue MySQL server to steal sensitive files or intercepts and downgrades secure traffic to redirect data to attacker-controlled endpoints.
\textbf{\textit{(4)}} \tool uncovered a compound squatting attack on \texttt{@typescript-eslint/eslint-plugin}, where an attacker used a similar namespace and package identifier (\texttt{@typescript\_eslinter/eslint}) to mislead users.

\begin{table}[ht]
\centering
\caption{
  \tool's detection counts by confusion‐attack category in production. 
  This total exceeds the numbers in \cref{tab:EQ5_Accuracy}, as it reflects the cumulative counts collected over the six-month production deployment.
  Traditional lexical confusion types (targeted by Typomind and SpellBound) are listed first, then additional semantic confusions escaping prior work, and finally new patterns (\cref{tab:typosquat_taxonomy}).
} 
\vspace{-0.1cm}
\label{tab:detection_counts}
\footnotesize
\begin{tabular}{p{0.4\columnwidth} >{\raggedleft\arraybackslash}p{0.3\columnwidth}|r}
\toprule
\textbf{Attack Category} & \textbf{Counts (Pct.)} & \textbf{Typomind} \\
\midrule
Traditional Lexical Confusion           & 431 (68.4\%) & 431 \\
\addlinespace
\midrule
Additional Semantic Confusion           &  69 (11.0\%) & 0 \\
Impersonation Squatting                 &  78 (12.4\%) & 0 \\
Compound Squatting                      &  50 ( 7.9\%) & 0 \\
Domain Confusion                        &   2 ( 0.3\%) & 0 \\
\midrule
Total                                   & \NumReviewedThreats (100\%) & 431 (68\%) \\ 
\bottomrule
\end{tabular}
\vspace{-5mm}
\end{table}

\subsection{EQ5: System Accuracy and Efficiency}
\label{sec:EQ5}

\ifNOBOXES
\begin{tcolorbox} [width=\linewidth, colback=yellow!30!white, top=1pt, bottom=1pt, left=2pt, right=2pt]
\textbf{EQ5 Summary:}
Our system exhibits slower processing speeds due to additional sorting and LLM verification steps but maintains acceptable efficiency and strong scalability for deployment on SPRs.
\end{tcolorbox}
\fi

\myparagraph{Method}
We evaluated the end-to-end system accuracy and efficiency  (\ie latency and throughput) by running \tool and \texttt{Typomind} on  \texttt{NeupaneDB}
and \texttt{ConfuDB}.
We also performed a top-$k$ analysis for \tool, varying the number of nearest neighbors in the benignity check to assess its impact on performance.


\myparagraph{Result}
\cref{tab:EQ5_Accuracy} presents the accuracy measurements.
Across the datasets, \tool generally exhibits stronger performance in detecting active threats, achieving higher recall than \texttt{Typomind} while both maintain perfect precision. In the case of stealthy threats, the results are mixed: \texttt{Typomind} excels in the \texttt{NeupaneDB} dataset with a perfect recall, yet its relative performance declines on \texttt{ConfuDB}. For benign cases, the accuracy differs notably, as \texttt{Typomind} misclassifies benign packages in \texttt{NeupaneDB}, whereas it performs better on \texttt{ConfuDB}, though still lower than \tool's performance in that category. Overall, \tool demonstrates strengths in detecting active and stealthy threats across datasets, reducing the FPR from \TypomindFPRate (\cref{sec:ProblemState-CaseStudy2-FP}) to \ConfuguardFPRate (133/\FPBenignPkgs).

Our efficiency measurement shows that \tool incurs much higher latency (\(31s\) per package \vs \(17s\) for Typomind) and correspondingly lower throughput (0.56 \vs 1.76 packages/\(s\)).
However, in the broader deployment context, \tool significantly reduces the workload on analysts, resulting in notable end-to-end cost savings.
The additional latency, stemming from benignity filtering and reliance on LLM-based benignity check, substantially enhances detection accuracy (reducing false positives).

Our top-$k$ analysis shows that \(k=2\) offers the best balance.
Moving from \(k=1\) to \(k=2\) raises \tool's recall from 58\% to 64\%, accuracy from 54\% to 58\%, and F1 from 67\% to 71\%, while latency increases from about \(10s\) to \(14s\).
Increasing to \(k=3\) or 4 yields gains of less than 2 percentage points in each metric but pushes latency to \(17s\) and \(21s\), respectively.
Our partner's separate malware scanner is expected to capture the remaining cases.

\section{Discussion} \label{sec:discussion}

\subsection{Improvements on Prior Detectors} \label{sec:design_advances}


Having described \tool's design in~\cref{sec:SystemDesign} and quantified its performance in~\cref{sec:Eval}, we now summarize \tool's four key advances over previous detectors such as Typomind and SpellBound:
\begin{enumerate}[leftmargin=*]
    \item \textbf{Hierarchy-aware detection (\cref{sec:SystemDesign-Step3}).} By splitting and scoring each namespace component, \tool can detect compound and impersonation squats in hierarchical naming ecosystems.
    Previous detectors only treated flat naming ecosystems, reducing their ability to identify the attack strategies of 
      \textit{impersonation squatting},
      \textit{compound squatting},
      and
      \textit{domain confusion} (\cref{sec:EQ4}).
    
  \item \textbf{Syntactic-Semantic Matching at Scale (\cref{sec:SystemDesign-Step4}).} 
  \tool leverages a fine-tuned \texttt{FastText} model to identify potential package confusions with high recall and low latency, avoiding the exhaustive comparisons typical of prior tools (\cref{sec:EQ1}, \cref{sec:EQ2}).

  \item \textbf{Metadata-driven benignity filter (\cref{sec:SystemDesign-Step5}).} 
  \tool incorporates 15 heuristics based on package metadata to accurately identify package confusions.
  This technique reduces the false positive rate by \FPReducedRate, compared to prior approaches (\cref{sec:EQ3}, \cref{sec:EQ4}).
  
  \item \textbf{Engineered for production deployment (\cref{sec:SystemDesign-Step6}).} 
  \tool can be integrated into existing security workflows, such as those used by our industry partner.
  It detects package confusion threats at the scale required by modern software package repositories --- which prior tools cannot achieve (\cref{sec:EQ5}). 
  In practice, \tool reduces manual review effort and cost for security analysts by over 50\%, so that even with higher per-package processing time, operational efficiency improves substantially.
\end{enumerate}
\vspace{-0.1cm}

\subsection{Limitations}
\label{sec:Discussion-LimitationAndSecurityAnalysis}
\label{sec:SystemDesign-Limitation}


The primary design limitation of \tool is its reliance on software metrics to gauge the likelihood that a package is a confusion attack.
These metrics might be gamed.
There has been little formal study of this risk, but recent work suggests both the possibility and some real-world examples~\cite{he2024FakeStars, kalu2025arms}.

A second design limitation is that the neighbor search relies on sub-word embeddings, which struggle to distinguish acronyms and short names (curse of dimensionality).
\textit{FastText} struggles with short words (\eg \texttt{xml} \vs \texttt{yml} have a similarity score of 0.7).
To mitigate attacks on short package names,
we implemented a list of potential substitutions to identify cases of visual or phonetic ambiguity, and we
combine embedding similarity with Levenshtein distance.



In our implementation, we rely on an LLM for benign filtering.
This introduces correctness risks (hallucination, jailbreaking~\cite{yi2024jailbreak}).
Adversaries might exploit this by tailoring their package name or metadata to persuade the LLM that the package is trustworthy using techniques like prompt injection or model hijacking~\cite{liu2023promptInjectionAttack, zou2023universalandTransferableAdvAttackonLM}. 


\subsection{Lessons Learned from Production} \label{subsec:three-lessons}
We share four lessons learned over the six months of production deployment of \tool.

\myparagraph{Lesson 1: \tool prevents real confusion attacks.}
Over the past six months, we deployed the system in our industry partner's production environment, during which we identified and confirmed \NumProdReviewedTPTypos threats, with \NumProdUnreviewedTypos additional threats under review.
Based on the false positive rate in our analysis so far, we estimate that among these threats, there are $\sim$17,500 real attacks.

\myparagraph{Lesson 2: Simplified LLM Prompts Unlock More Robust Reasoning.}
\\ 
Our production experience revealed that overly detailed, rule-by-rule prompts can constrain an LLM’s ability to synthesize diverse signals, leading to both blind spots and bias~\cite{gu2025surveyllmasajudge}.  For example, by instructing the model to treat any package with a namespace as “likely fork,” we missed malicious variants that reused legitimate namespaces but embedded suspicious payloads in their README or metadata.  We replaced the myriad micro-rules with a decision framework approach (\cref{lst:llm-prompt}), so the LLM can weigh name similarity, maintainer reputation, and other signals together.

\myparagraph{Lesson 3: Ontology Matters.}
\tool is deployed as part of a security analysis pipeline, with human analysts to triage the results (\cref{sec:SystemDesign-Step6}).
Analysts must interpret the results of \tool,
and feel the package confusion categories outlined in \cref{tab:typosquat_taxonomy} do not provide sufficient explanatory power.
We propose to refine the confusion taxonomy to consider the malicious content and intent behind each package, including a risk-level classification.

\myparagraph{Lesson 4: Large-Scale Empirical Feedback is Crucial.}
Our work with \tool and the experience-driven optimization pipeline (\cref{sec:SystemDesign-Step6}) underscores that only through large-scale empirical measurement can we systematically uncover the bottlenecks, failure modes, and hard edge cases that theory alone cannot predict. By continuously injecting human-in-the-loop analysis --- where experts inspect anomalies and feed targeted insights back into the design --- we close the gap between observed behavior and ideal performance by turning raw operational data into actionable system improvements.

\myparagraph{Lesson 5: False Positives Harm Our Customers.}
\cref{tab:metadata_rules} represents the most recent benignity check rules. In deployment, we found additional cases where a package had a very similar name and no README, which made our system classify it as a suspicious stealth attack.
That package was owned by one of our partner's customers and served as a ``transitive package''. 
The customer felt that being flagged harmed their reputation. 
To address such cases, where engineering convenience conflicts with  security~\cite{gopalakrishna__nodate} (cf.~\cref{sec:PracticalGaps}), we added an allow-list of our partner's customers. 

\subsection{Future Directions}
\label{sec:Discussion-FutureWork}

\myparagraph{Enhancing Representations.}
\label{sec:Discussion-FutureWork-Representation}
Improving the representation of package names is crucial for more robust detection. While FastText captures semantic similarity through subword embeddings, it struggles with typographical variations, particularly for short words or acronyms. 
In our experiments on Hugging Face packages, we observed uniformly high similarity scores due to long, acronym-dense names and extensive forking activity, which undermines standard confusion-detection methods~\cite{jiang2023PTMNaming}.
LLMs can generate rich, domain-specific corpora, complete with synthetic typos and acronym variations.
Applying (semi-)supervised learning to such tailored datasets can address these limitations.

One challenge based on our production experience is the limited availability of data for verifying package confusions.
LLMs might help by augmenting training data with synthetic typos~\cite{tozzi2024packageHallucinatinoThreat}. 
Advanced models, such as transformer-based architectures fine-tuned with contrastive learning on such typo-specific datasets, present a promising way for better detection accuracy and reliability.
This approach has proven effective in combating domain typosquatting, but no research has been conducted targeting package confusions~\cite{koide2023phishreplicantLanguageModelBasedDNSSquattingDetection}. 

\myparagraph{Human-in-the-Loop Optimizations.} 
Our deployed system offers opportunities for further optimization, particularly through the incorporation of additional human-annotated ground truth. Drawing inspiration from human-in-the-loop reinforcement learning --- where domain experts iteratively guide and refine agent behaviors --- and self-adaptive prompting~\cite{wan-etal-2023-better, retzlaff2024HILRL}, we propose integrating real-time feedback from security analysts into the reasoning processes of LLMs~\cite{gerosa2024canAI}. Analysts' corrections and augmentations of intermediate model outputs serve as dynamic expert feedback, enabling improvements in both the prompts and the model's behavior.

\myparagraph{Anomaly Detection for Malicious Package Detection.}
Anomaly detection approaches offer significant potential for improving malicious package detection.
By leveraging anomaly detection techniques and metadata analysis~\cite{halder2024maliciousPackageDetectionUsingMetadataInfo}, future package confusion detectors might dynamically adapt to evolving attack strategies, \eg learning new benignity rules without human oversight. 

\section{Conclusion}

Modern software applications integrate third-party packages.
Attackers therefore publish packages with names deliberately intended to confuse engineers, seeking to inject malware into production systems.
Accurately detecting these attacks at ecosystem scale requires a low-latency system that accounts for human interpretations of package names. 
We showed that these requirements can be met by an embedding-based detection system combined with heuristics based on real-world attack patterns. 
Compared to the state-of-the-art, our \tool system
  captures additional confusion categories,
  achieves a substantially lower false-positive rate,
  and maintains acceptable latency for use either on registry backends or in handling on-demand requests.
We conclude with insights from production experience and customer feedback. 

\section*{Data Availability}
\label{sec:OpenScience}
Our artifact is available at:
\url{https://github.com/confuguard/confuguard}.
It includes everything but the commercial metadata database. 

\JD{Fix many warnings/errors so we don't miss broken cites like in \$7.3.}

\pagebreak

{
\normalsize
\bibliographystyle{plain}
\bibliography{main.bib}

}



\clearpage

\end{document}